\documentclass[twocolumn,showpacs,preprintnumbers,amsmath,amssymb,prb]{revtex4}

\usepackage{graphicx}
\usepackage{dcolumn}
\usepackage{bm}

\newcommand{\etal}{{\it et al.\ }}

\begin{document}

\preprint{APS/123-QED}

\title{Classical electrons in laterally coupled diatomic 2D artificial
molecule}

\author{M. Marlo}
 \email{Meri.Marlo@hut.fi}
\author{M. Alatalo}
\author{A. Harju}
\author{R. M. Nieminen}
\affiliation{
Laboratory of Physics, Helsinki University of Technology, P. O. Box
1100 FIN-02015 HUT, Finland   
}

\date{\today}

\begin{abstract}
Structural properties of a finite number ($N = 2 - 20$) of 
point charges (classical electrons)
confined laterally in a two-dimensional two-minima potential are
calculated as a function of the distance ($d$) between the minima. The
particles 
are confined by identical parabolic potentials and repel each other
through a Coulomb potential.  Both ground state and metastable
electron configurations are discussed. At zero distance previous 
results of other calculations and experiments are reproduced. 
Discontinuous transitions 
from one configuration to another as a function of $d$ are
observed for $N = 6, 8, 11, 16, 17, 18, 19$. 
\end{abstract}

\maketitle

\date{\today}

\maketitle

\section{INTRODUCTION}
\label{sec:intro}

Quantum dots (sometimes called artificial atoms) are
nanoscale semiconductor structures where a small 
number of electrons are confined into a small spatial
region.\cite{Ashoori,Kastner} The electron motion is usually
further restricted to two dimensions.       
There is strong theoretical evidence for the existence
of a limit where the electron system 
crystallises to Wigner
molecules, which 
is seen as the localisation of
the electron density 
around positions that minimise
the Coulomb repulsion.~\cite{Egger,Reimann,Creffield,Reusch,Arin_paperi,Koskinen,Maksym_molaspects}         
In the limit of weak confinement (low density) or a very strong
magnetic field the quantum effects are quenched or obscured and
the classical electron correlations start to dominate the
properties of the system. 
The ultimate limit is a purely classical system where only the
Coulomb repulsion between the electrons defines the ground state.
The problem reduces to finding the classical positions of
electrons (which depend on the forms of the confining
and the interaction potentials) that minimise the total energy of the
system.   

There is growing interest in
calculating\cite{Imamura,Nagaraja,Partoens,Veikon_paperi,Wensauer,Kolehmainen,Jannouleas}  
and measuring\cite{Livermore,Oosterkamp,Brodsky}  the
properties of coupled quantum
dots.
Due to the 2D nature of
quantum 
dots the two-atom system is different 
whether the quantum dots are coupled in the plane 
in which the electrons are confined ({\it laterally} coupled)
or in the perpendicular direction ({\it vertically}
coupled). Especially for {\it laterally} coupled quantum dots
only a limited number of studies have appeared.\cite{Veikon_paperi,Wensauer,Kolehmainen,Jannouleas}  
Classical studies serve as a good starting  
point for more demanding quantum mechanical calculations. Moreover,
the study of classical electrons in {\it vertically} coupled 
artificial atoms has revealed interesting structural transitions in the
ground state electron configurations as a function of the  
distance between the atoms.\cite{PartoensClassicalPos}

Apart from quantum dots in the classical limit the point charges in 2D
can be used to model also other physical systems.
Examples include
vortex lines in superconductors and superfluids and electrons on the
surface of liquid He (see Ref. \onlinecite{Kong} and
references therein). In the theoretical field,
the ground state configurations of a confined classical 2D electron
system have been studied in the case of a single artificial atom in 
Refs.
\onlinecite{Kong,Bolton_Superlatt,BedanovPeeters,Ying-Ju,Date,Campbell,Schweigert}
and for the vertically coupled artificial atom
molecule as a function of the inter-atom distance in
Ref. \onlinecite{PartoensClassicalPos}.  
Recently, also some experimental studies of 2D confined charged classical
particle systems have appeared to reflect the classical cluster patterns
in 2D.\cite{dustparticles,exp_classical}

 Classical point charges in a two-dimensional
 infinite plane crystallise into a hexagonal lattice at low
 temperatures. Parabolic 
 confinement in the artificial atom, on the other hand, favours
 circularly symmetric 
 solutions. The ground state configuration is thus determined by two
 competing effects, circular symmetry and hexagonal coordination, thus
 resulting in non-trivial particle configurations.  
 The reported configurations of the electron
 clusters in a single artificial atom do not all agree between different
 studies. The differences can be partly explained by the different forms of
 confinement and interaction potentials. However, when the number of
 particles, $N$, confined in the atom is one of the following $N =
 2-5,7,10,12,14,19$ all results are in agreement 
 while differences appear for $N = 6,8,9,11,13,15-18,20$ (for $N
 \leq 20$).      

 In this paper we consider two laterally coupled artificial atoms and
 classical electrons in the molecule. The changes in the ground state
 electron configurations are studied for $N = 2 - 20 $ electrons in
 the molecule as the inter-atom distance is changed.  
 The energies of the 
 metastable states are also calculated at different distances and their
 role in the structural transitions in the ground state electron
 configurations is discussed. We also reproduce electron 
 configurations of the single parabolic artificial atom. The differences
 between different calculations and experimental results are discussed
 in the limit of single atom.     

\section{Monte Carlo simulation}

 The classical electrons in the artificial atom molecule are modelled
 with the Hamiltonian

\begin{eqnarray}
 H &=& \frac{1}{2} \ m^* \omega_0^2 \ \sum_{i=1}^{N} \min \left[ ({\vec
 r_i} -  d/2 )^2, ({\vec r_i}+d/2 )^2 \right]  \nonumber \\
 &+& \frac{e^2}{4 \pi \epsilon_0 \epsilon} \ \sum_{i<j}\frac{1}{|{\vec r_i}
 - {\vec r_j}|}.
\end{eqnarray}        
 Each of the $N$ electrons is described with coordinates $\vec{r_i} =
 (x_i, y_i)$ in two-dimensional space. 
 The harmonic confinements are positioned symmetrically around the
 origin with distance $d$ between the minima.  $m^*$
 is the electron effective mass, 
 $\omega_0$ the confinement strength and $\epsilon$ the dielectric
 constant. We measure the energy in meV and distance in
 \AA. The confinement strength was set to  $\hbar \omega_0 =
 3$ meV and typical GaAs parameters were chosen to the effective mass
 and the dielectric constant: $m^* = 0.067 \ m_e$ and 
 $\epsilon = 13$. The calculated energy values and distances could be
 scaled to correspond to different values of $\omega_0, \ m^*, \
 \epsilon$, but changing the parameters also changes the
 effective distance between the minima, $d$, and then the minimum
 energy configuration may not be the same anymore. Therefore we have
 have one significant parameter in the system, $d$, which scales as $
 \propto (m^* \omega_0^2 \epsilon)^{-1/3}$. 

 The minimum energy as a function of the positions of the particles, $
 E_{tot} = \min  
 \ E({\vec r_1},...,{\vec r_N})$, is solved with a standard 
 Metropolis Monte Carlo method~\cite{Metropolis} 
 starting from a randomly chosen initial electron configuration ${\vec
 r_1},...,{\vec r_N}$. The accuracy and simulation time needed with
 the Metropolis algorithm was found to be well sufficient for the
 current problem. We compared the calculated energies in the limit of
 a single artificial atom to those given in 
 Ref. \onlinecite{Kong}, and the results were found to be in complete
 agreement within the given accuracy. 

 In the simulations we choose four different distances between the
 atoms and perform 300 
 test runs at each particle number ($N = 2 - 20 $) and distance ($d =
 0, 200, 600$ and $1000$ \AA). In addition to minimum energy
 configurations we also obtain metastable states that are higher in
 energy compared to the ground state. 
 
 When the 
 ground and metastable states are obtained at $d = 0, 200, 600, 1000$
 \AA \  we study the structural transitions between ground state
 electron configurations at the intermediate
 distances.
 The electron configurations obtained
 from the fixed $d$ calculations are
 taken as an input to Monte Carlo minimisations where the attempt step
 is set so small that the electron configuration cannot change to
 another. Then the distance is slightly altered ($d \rightarrow
 d \pm 1$ \AA) and a new energy with slightly modified positions is
 calculated for the configuration defined by the input. 
 The calculated new configuration is taken as an input to the next
 calculation with a new distance between the
 atoms, and so all distances between $d = 0,200,600,1000$ \AA \ are
 well sampled. However, it may happen that a configuration becomes
 unstable as the 
 distance is changed. In that case the simulation converges to some other
 stable configuration, which can be seen as a sudden jump to a new
 energy value in the $E(d)$-plots. 
 The energies of all states are studied as a function of the distance
 and structural transitions in the ground state configurations are
 examined.

\section{Results}

The results for the ground and metastable states are summarised in Table
\ref{Tab:N}. The electron configurations are given at four different
distances between the artificial atoms. The ground state energy and
the corresponding configuration 
at the four distances is represented in the row following the particle
number $N$. 
If there exist  
metastable states at the given $N$ and $d$ the energy difference
$\Delta E/N$ to the ground state and the electron configuration for
the metastable state is also reported. 
However, not all metastable configurations are marked in Table \ref{Tab:N},
since when starting the simulation from random positions more
electrons can be 
trapped in one artificial atom than in the other. Only metastable states
with either the same number of electrons per atom (for even
$N$) or only one more at one than the other atom (for odd $N$) are
reported. 
The notation for the configurations in a single artificial atom is
chosen so that 
electrons are thought to be
organised in (nearly) concentric shells around the potential minimum:
(N$_1$,N$_2$,N$_3$), where N$_1$ denotes the number of electrons in
the innermost shell, N$_2$ the next shell and N$_3$ the number of
electrons in the outermost shell. (For $N \leq 20$ only three shells
are occupied). 
For laterally coupled two-atom artificial molecule we
have chosen the following notation for configurations:
At $d = 200$ \AA \ the configuration is marked as if it would still be on
a single atom centred around the midpoint connecting the two atoms. At
$d = 600$ and $1000$ \AA 
\ the configurations are given as configurations of two separated atoms.

\begin{figure}
$\begin{array}{|c|c|c|} \hline
\includegraphics*[height=25mm]{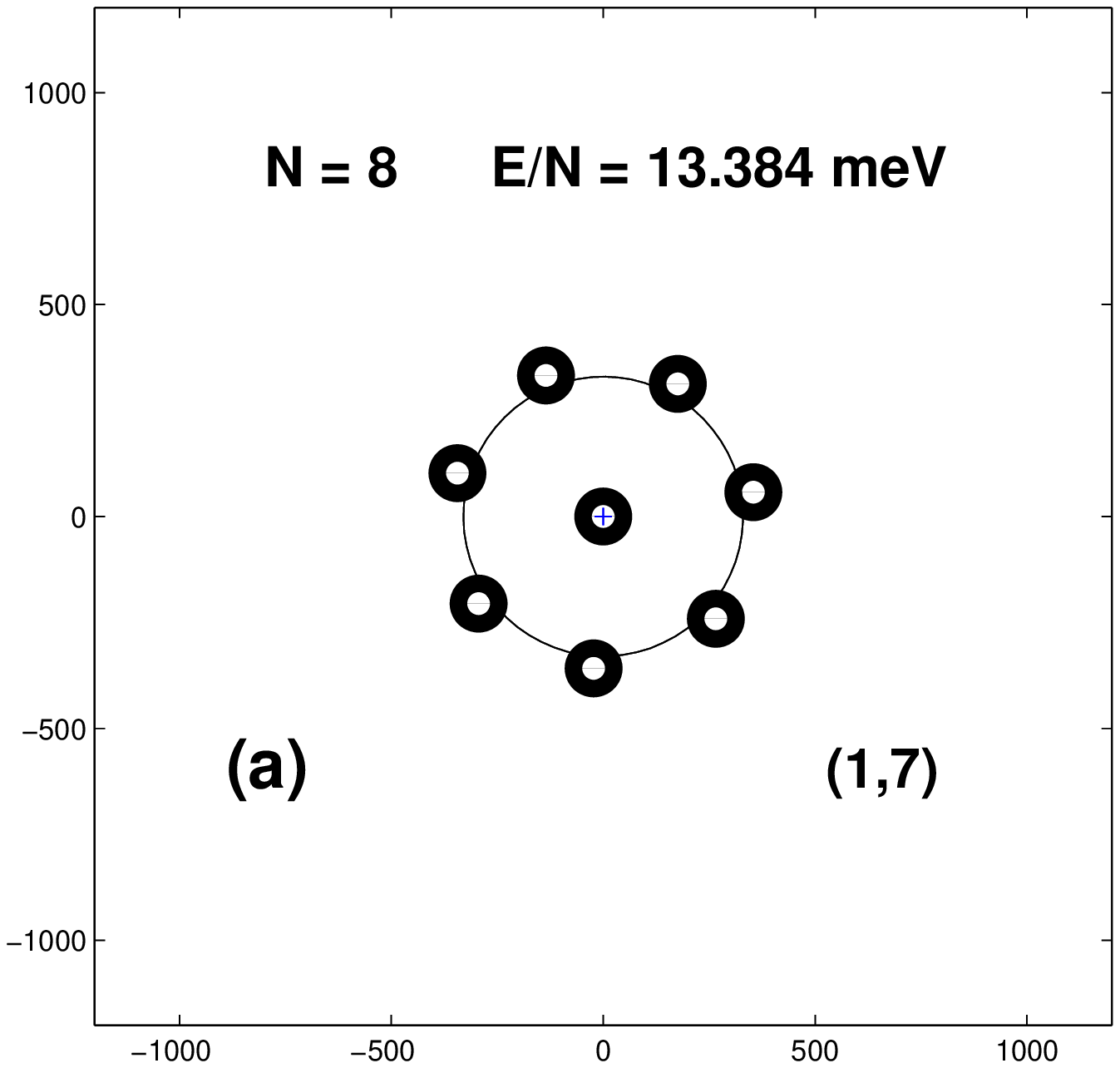} 
&  
\includegraphics*[height=25mm]{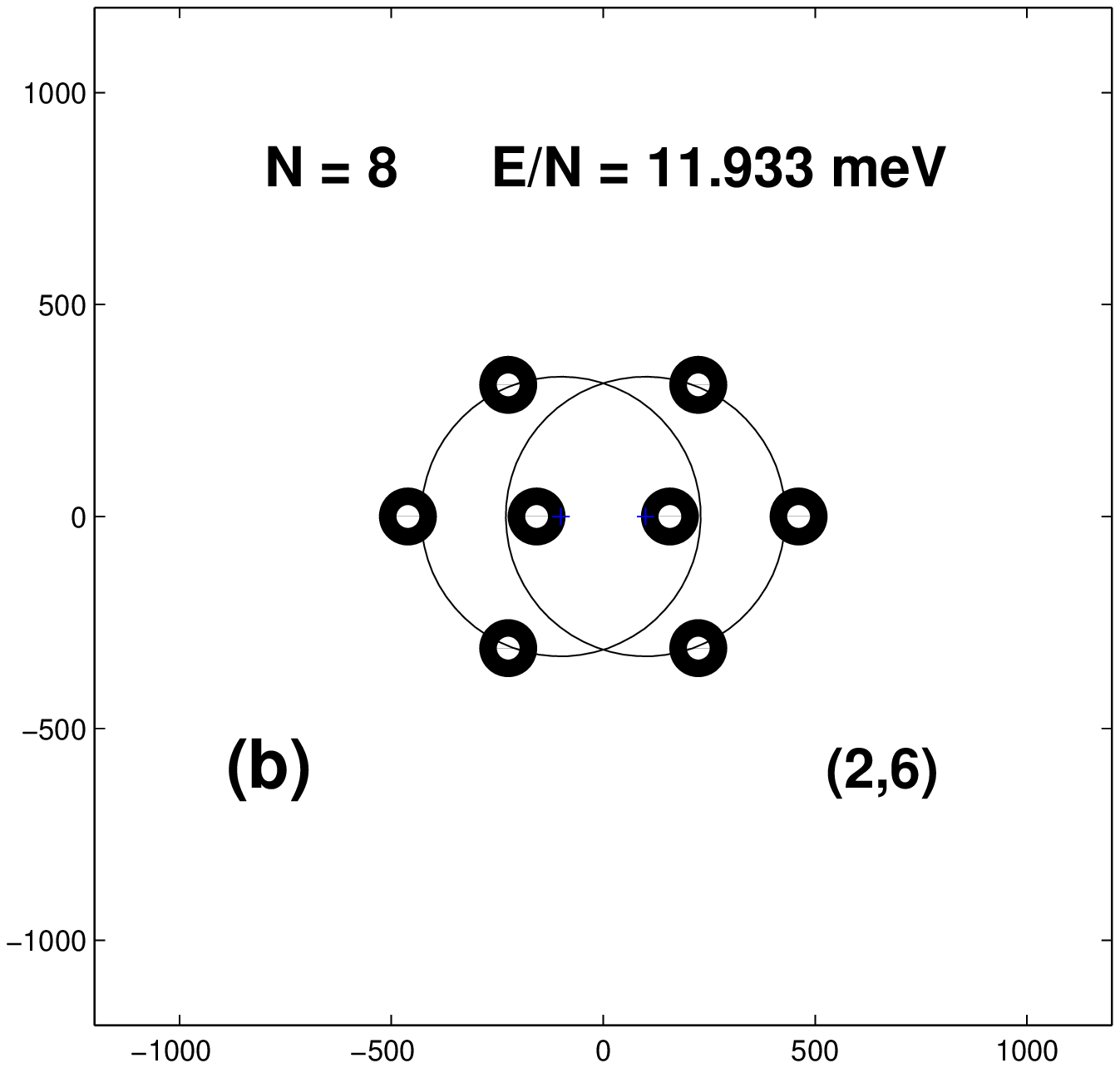}   
&  
\includegraphics*[height=25mm]{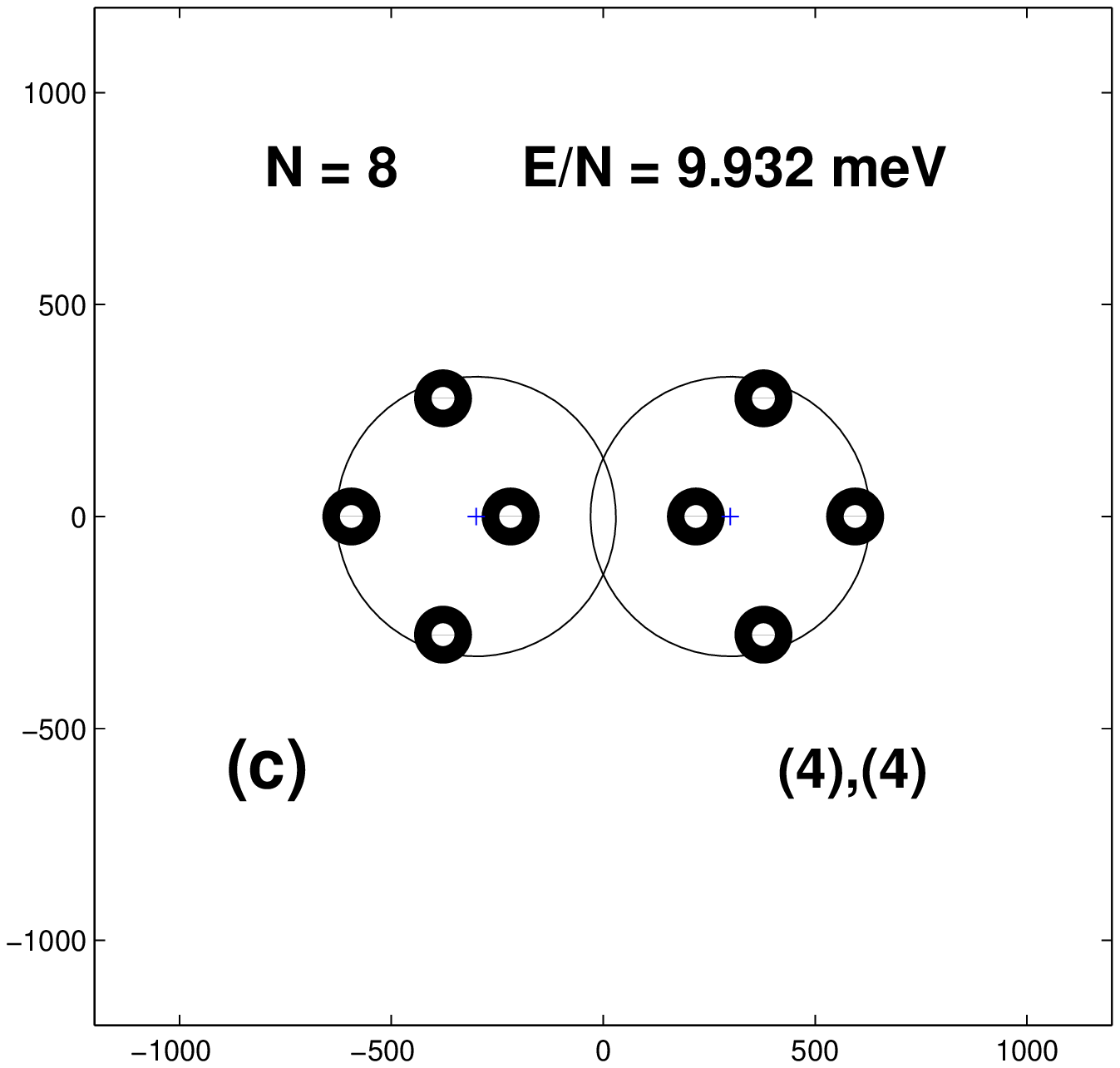} 
\\ \hline 

\includegraphics*[height=25mm]{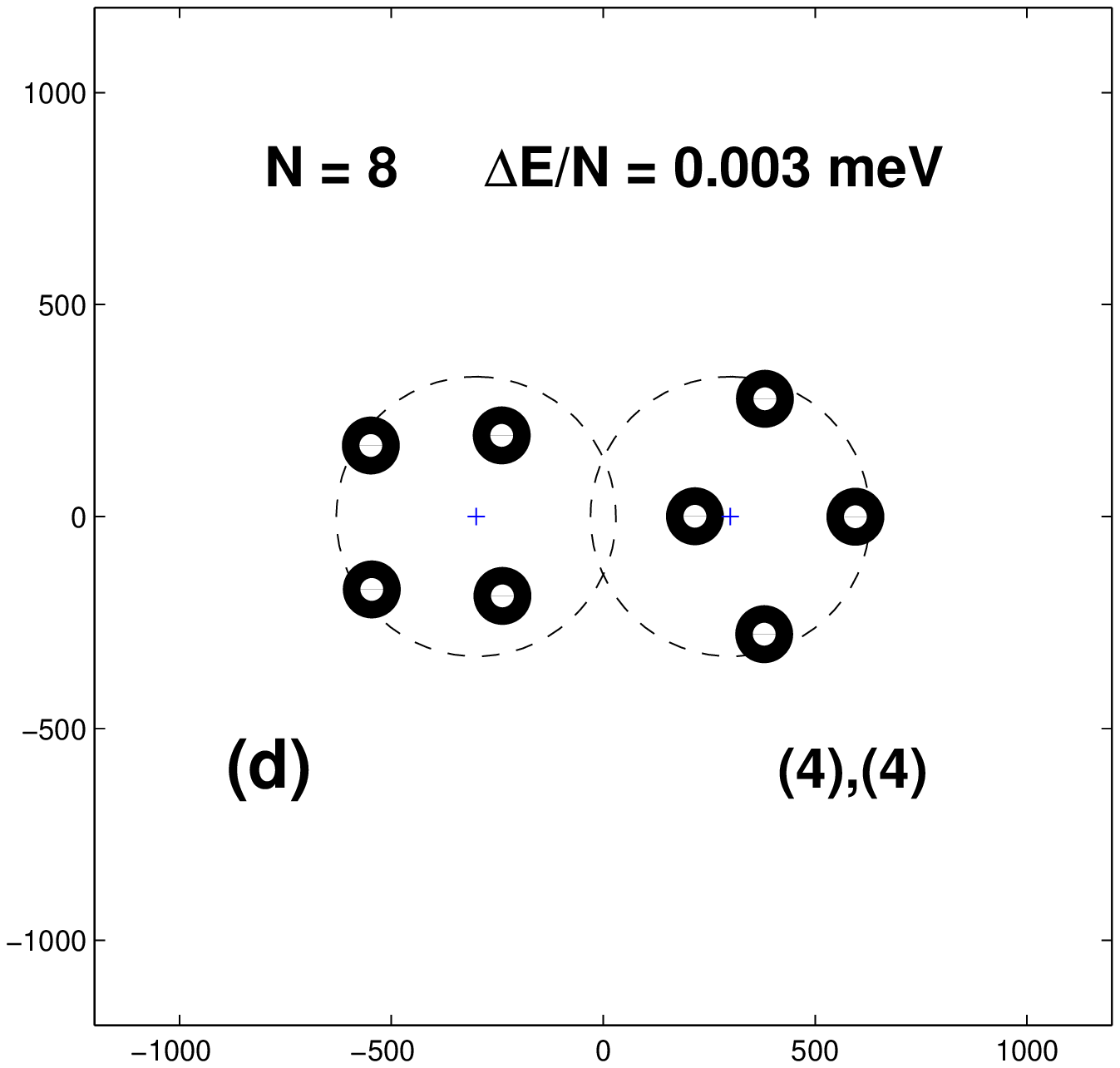} 
&
\includegraphics*[height=25mm]{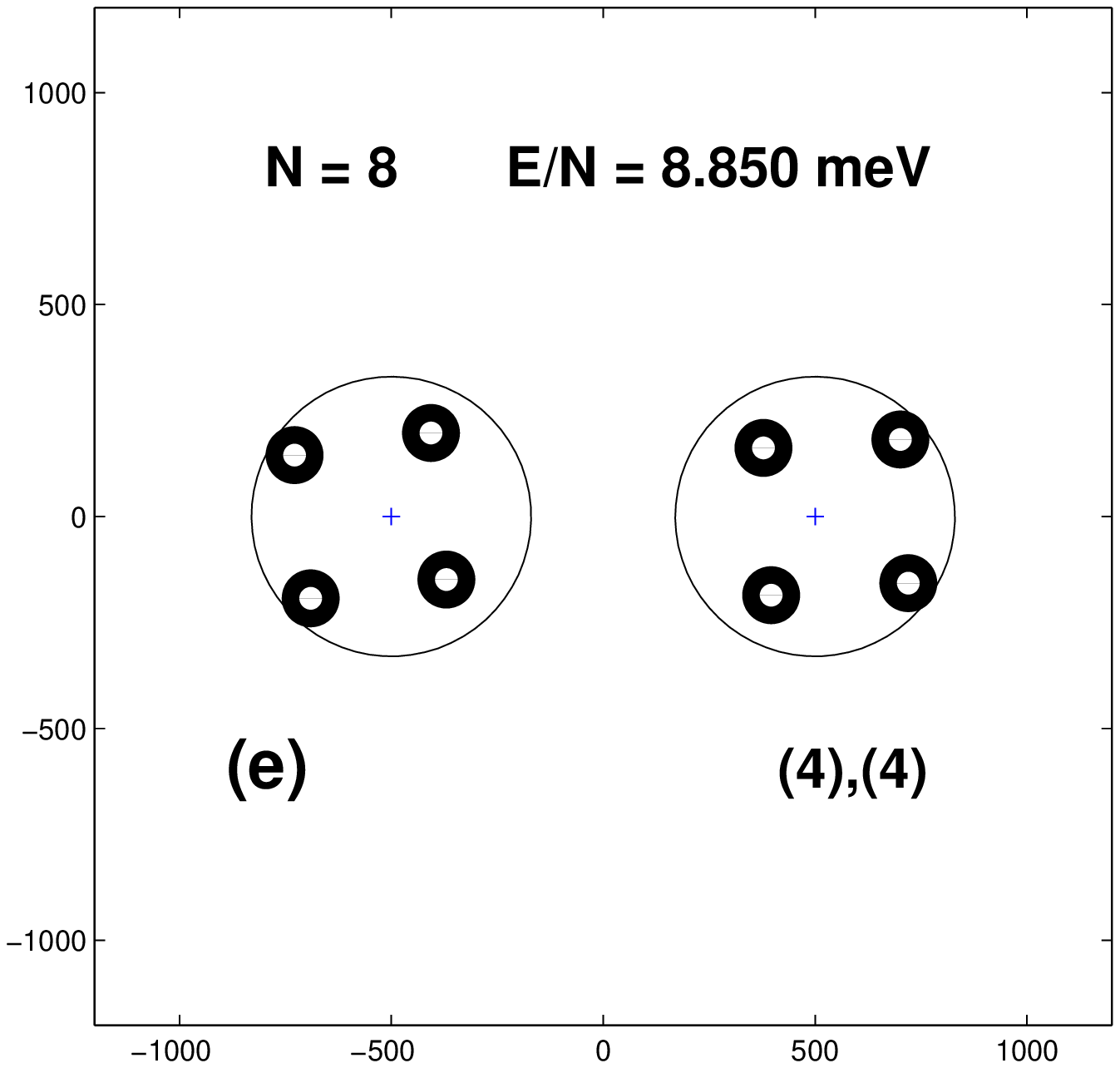} 
& 
\includegraphics*[height=25mm]{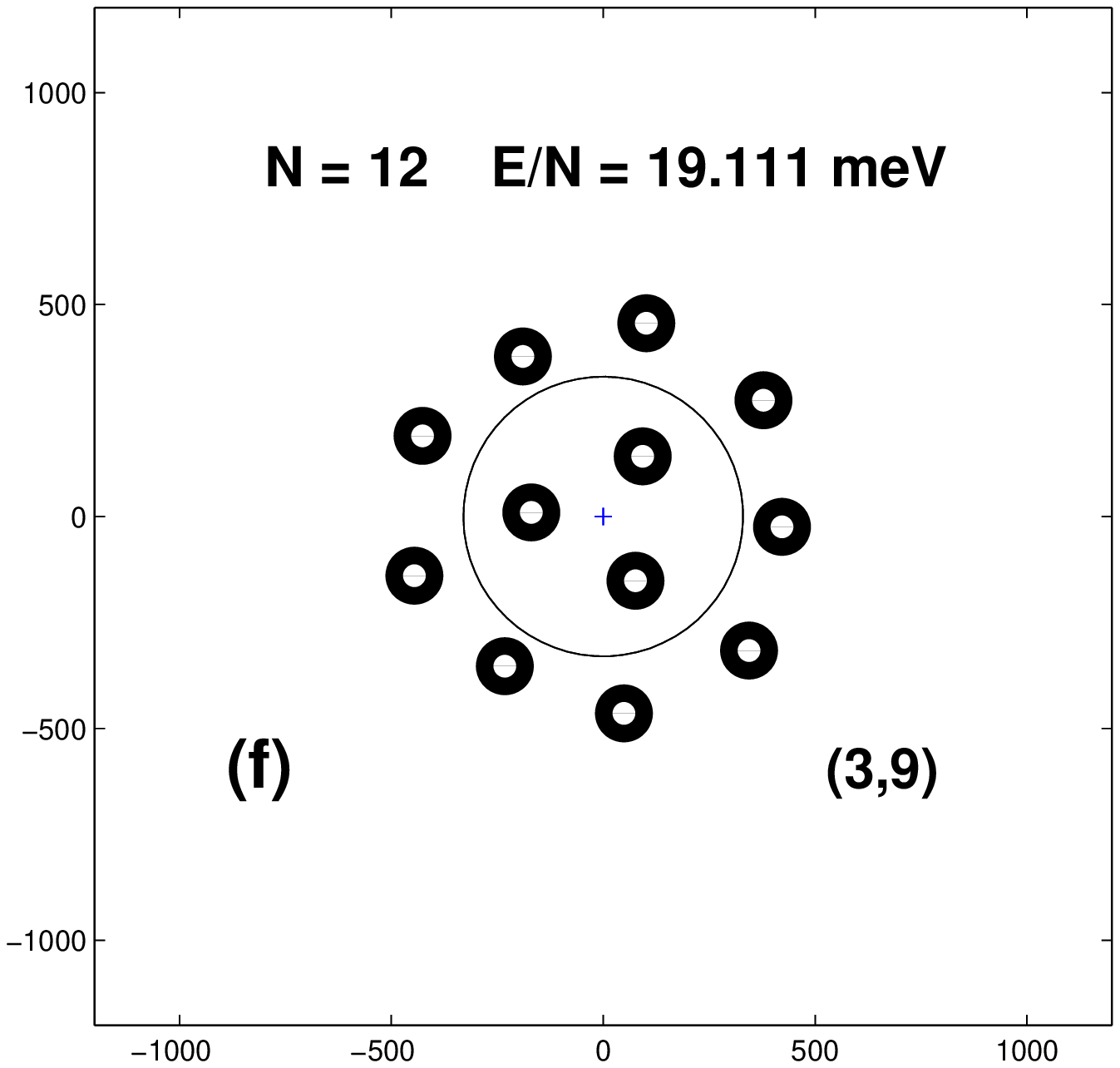} 
\\ \hline 

\includegraphics*[height=25mm]{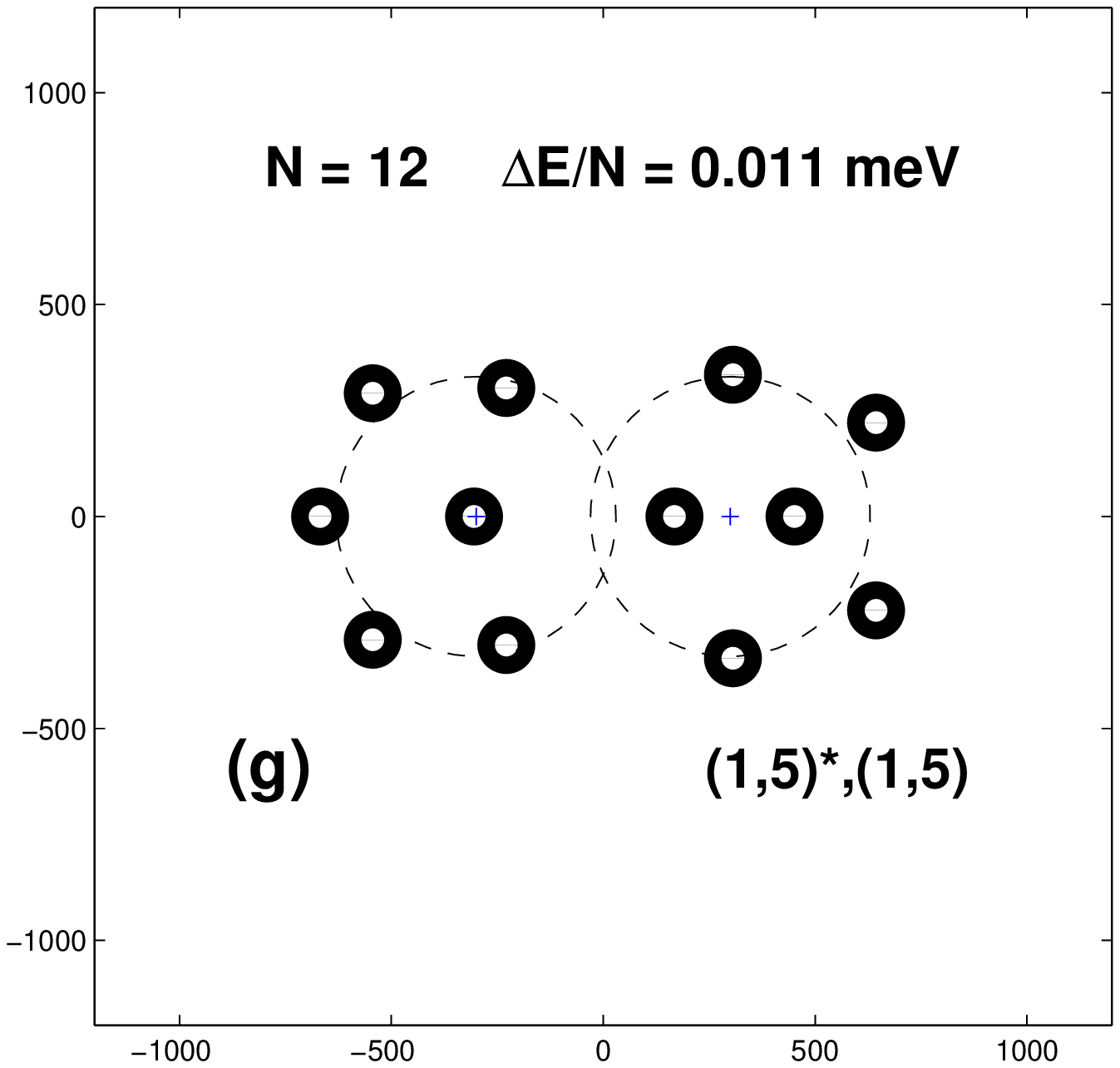}
&
\includegraphics*[height=25mm]{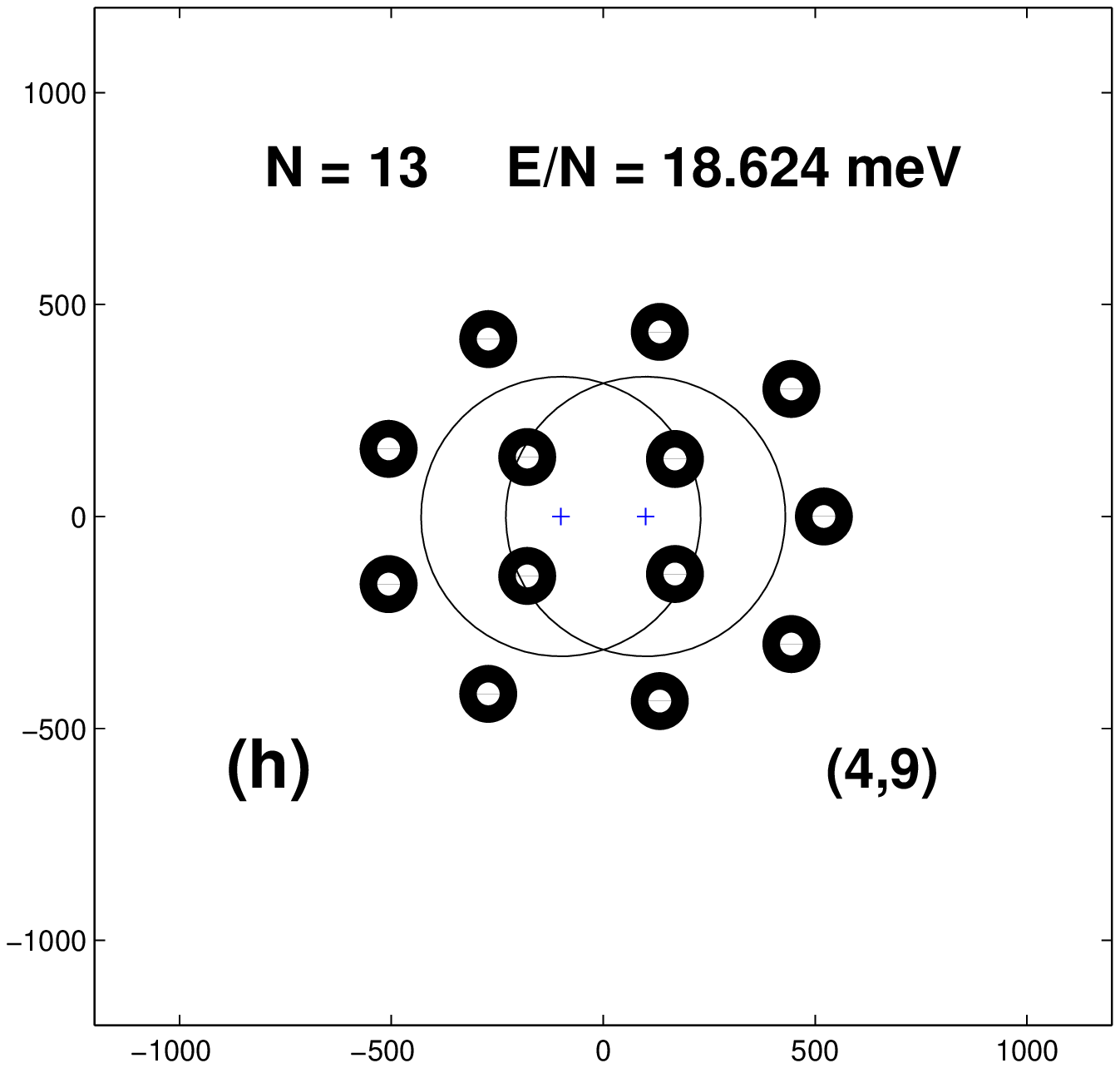} 
&    
\includegraphics*[height=25mm]{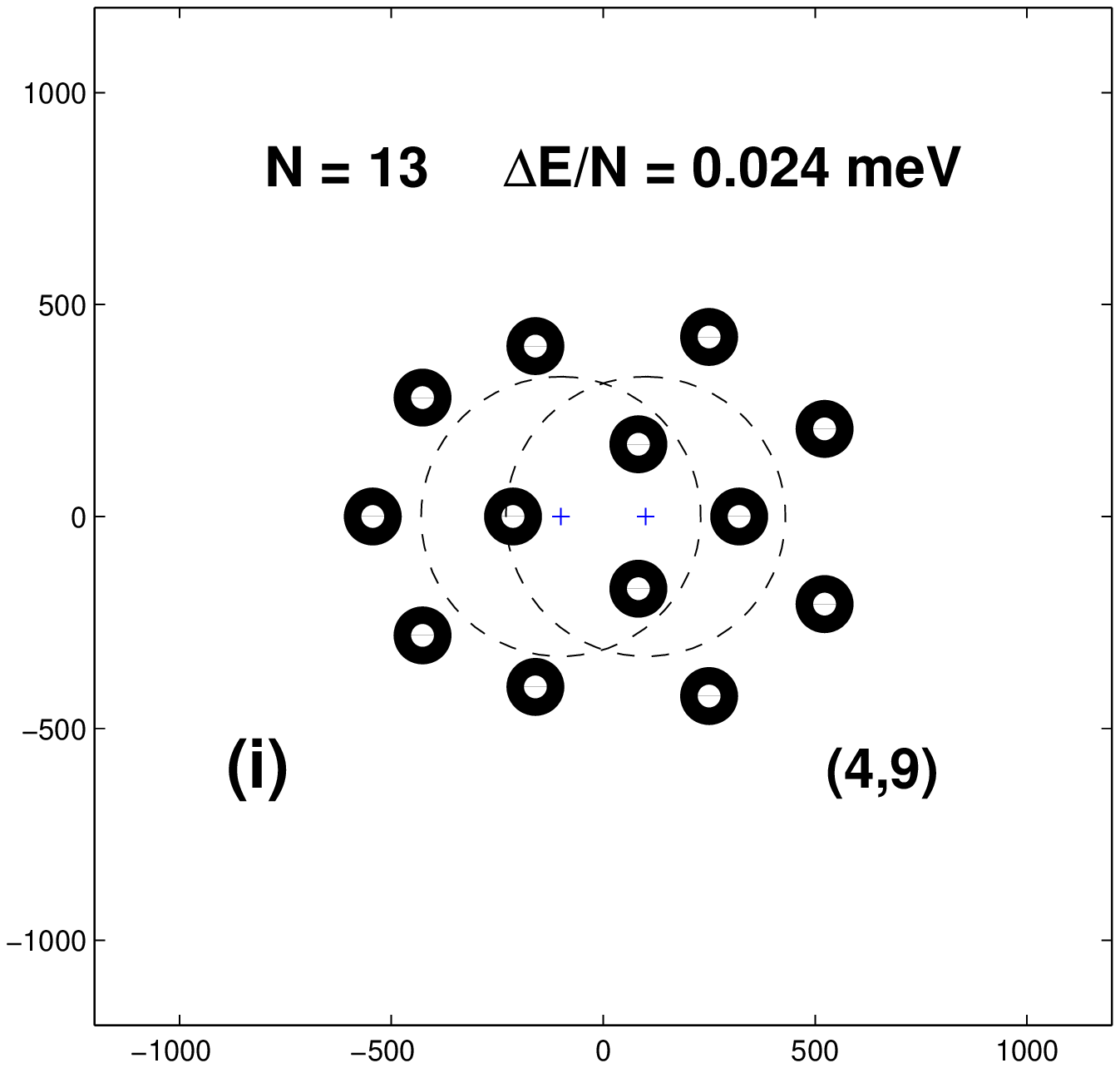} 
\\ \hline 

\includegraphics*[height=25mm]{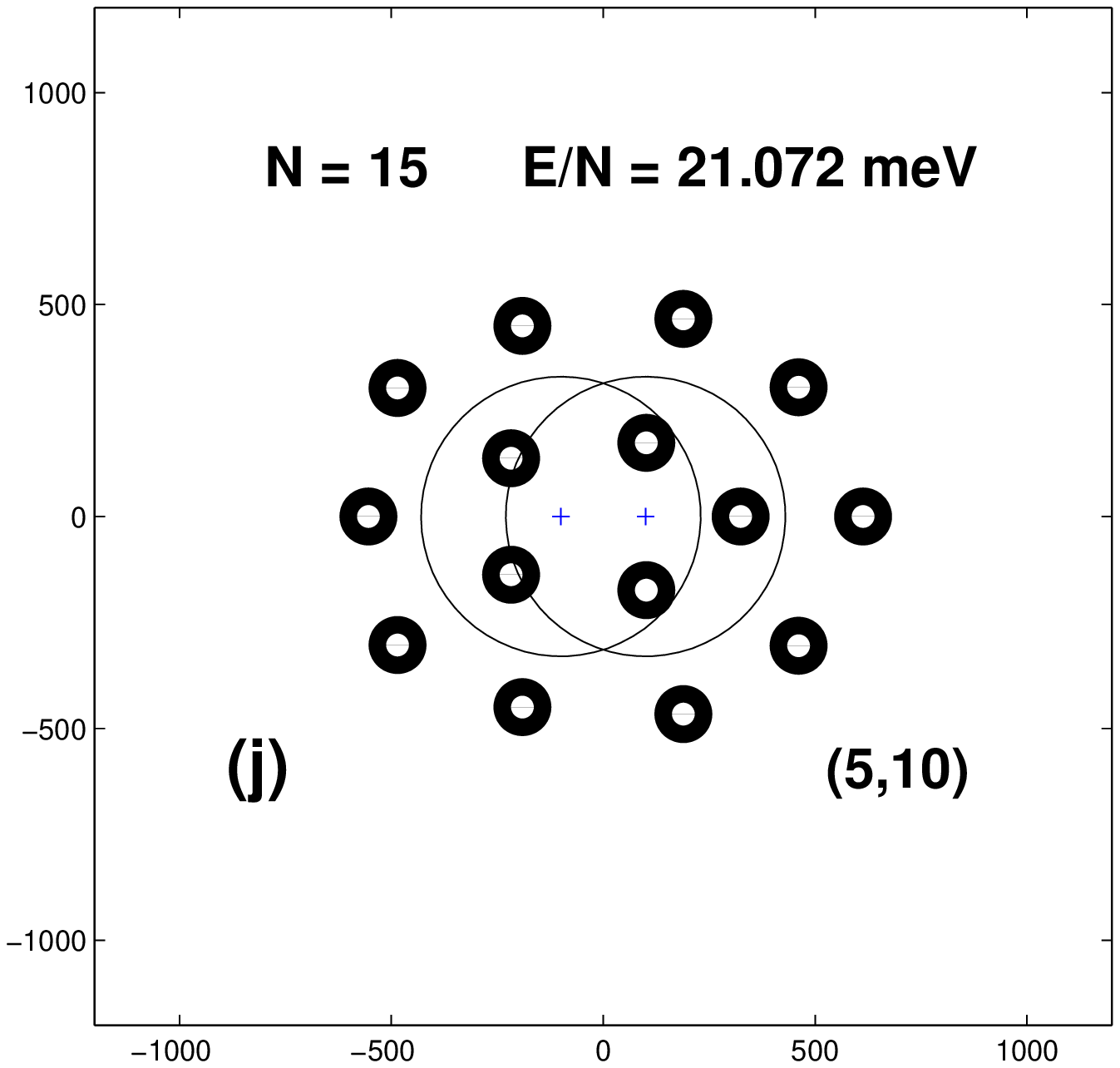} 
&
\includegraphics*[height=25mm]{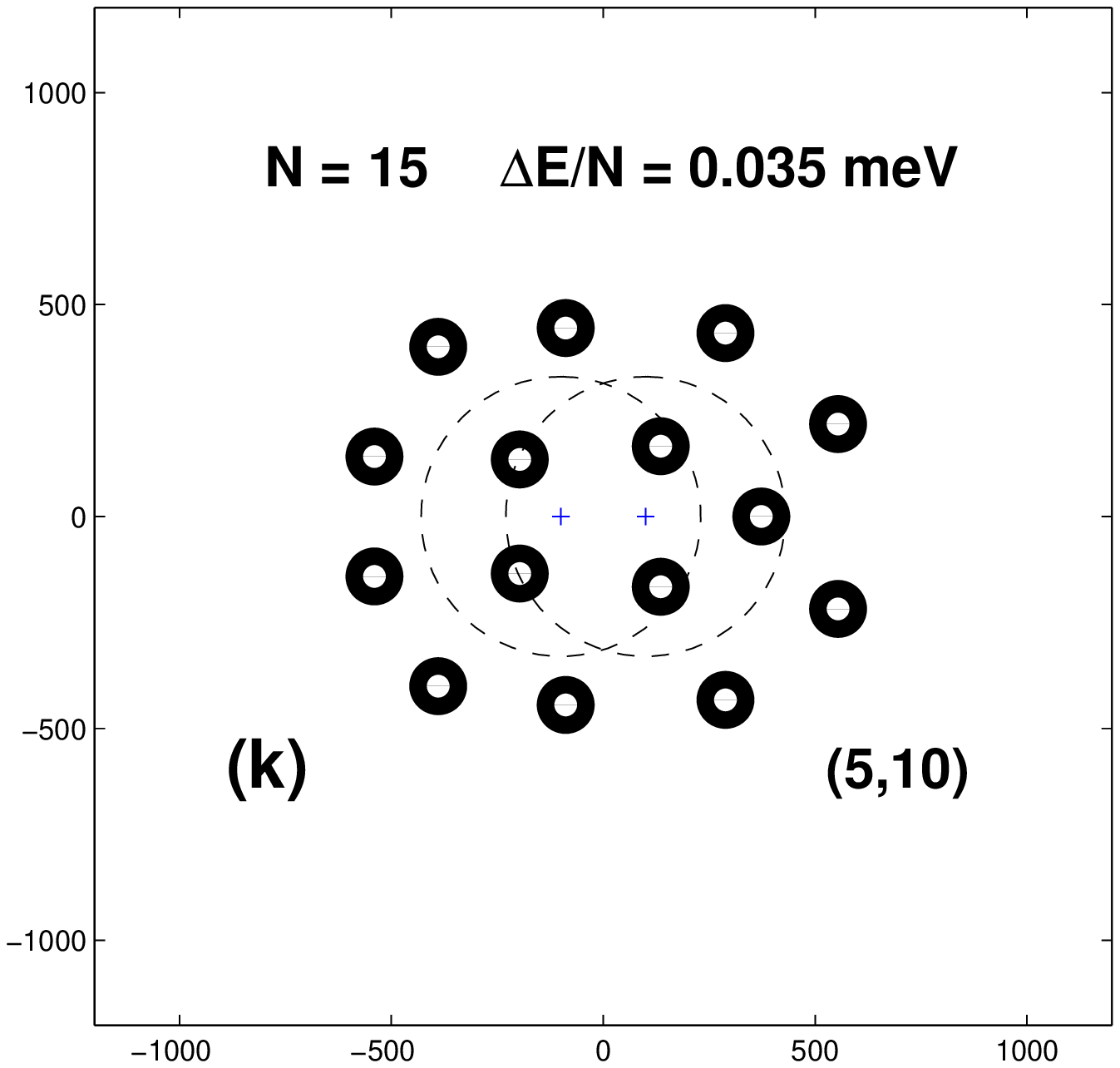}
&
\includegraphics*[height=25mm]{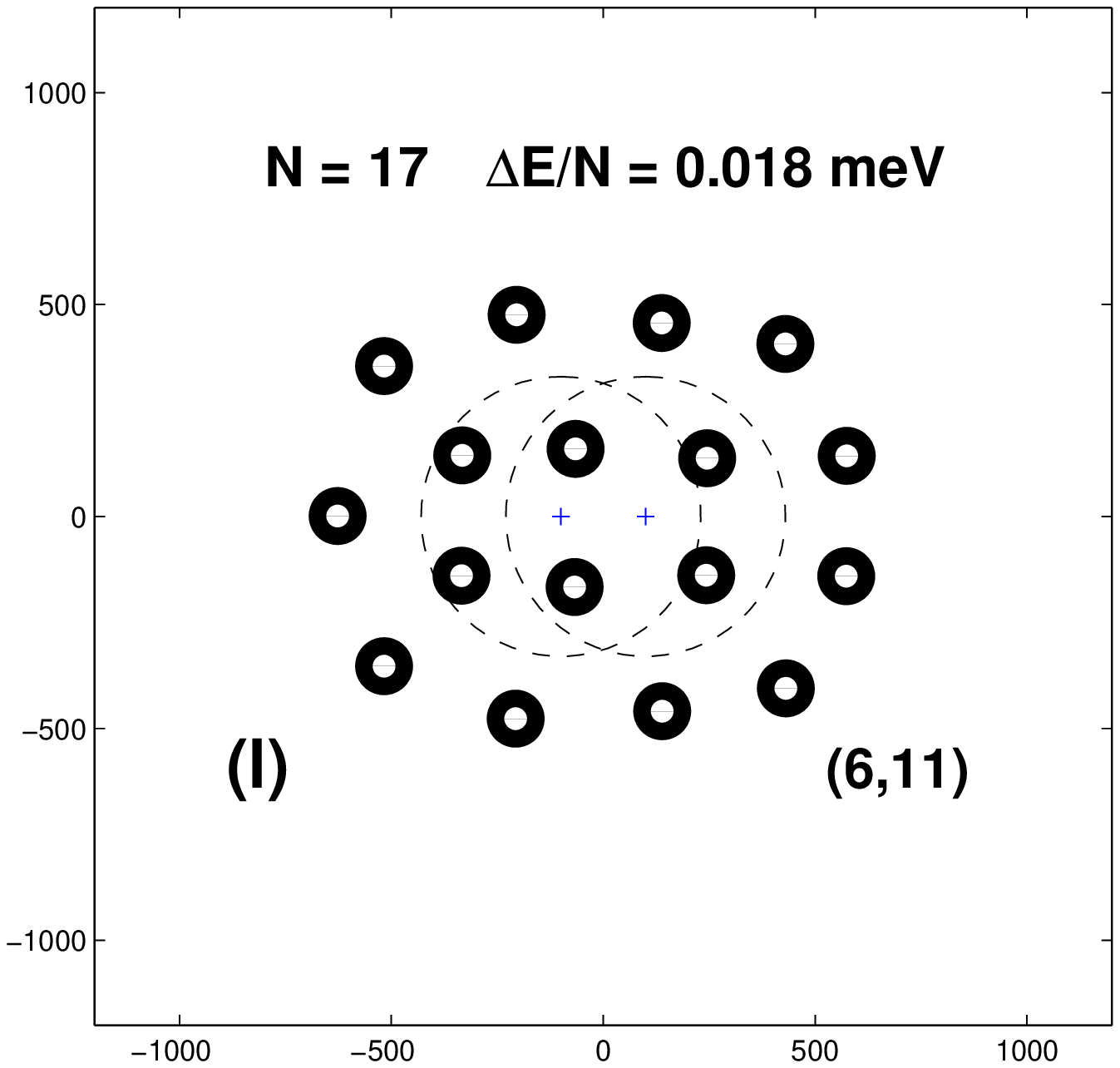} 
\\ \hline 

\includegraphics*[height=25mm]{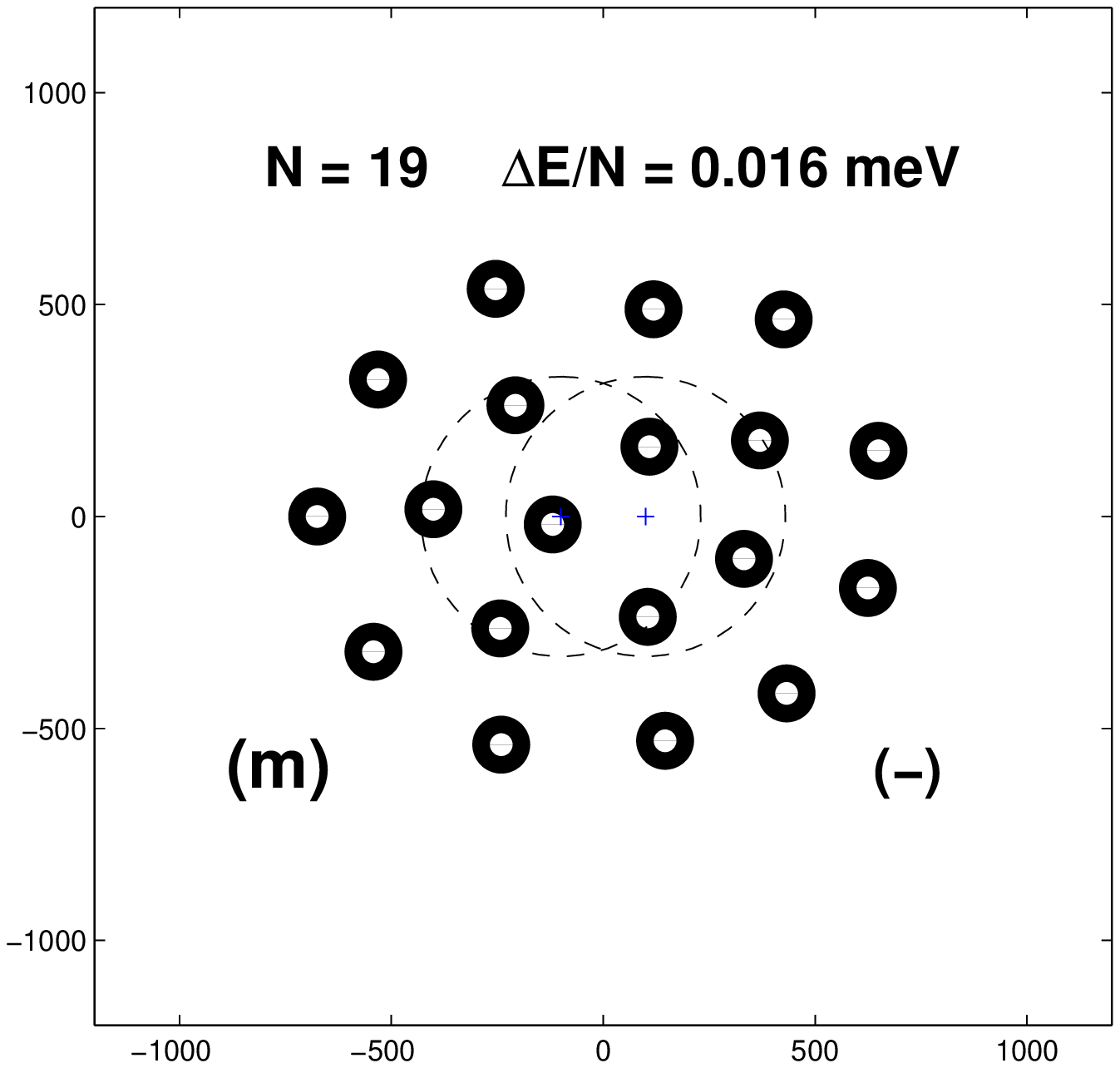} 
&
\includegraphics*[height=25mm]{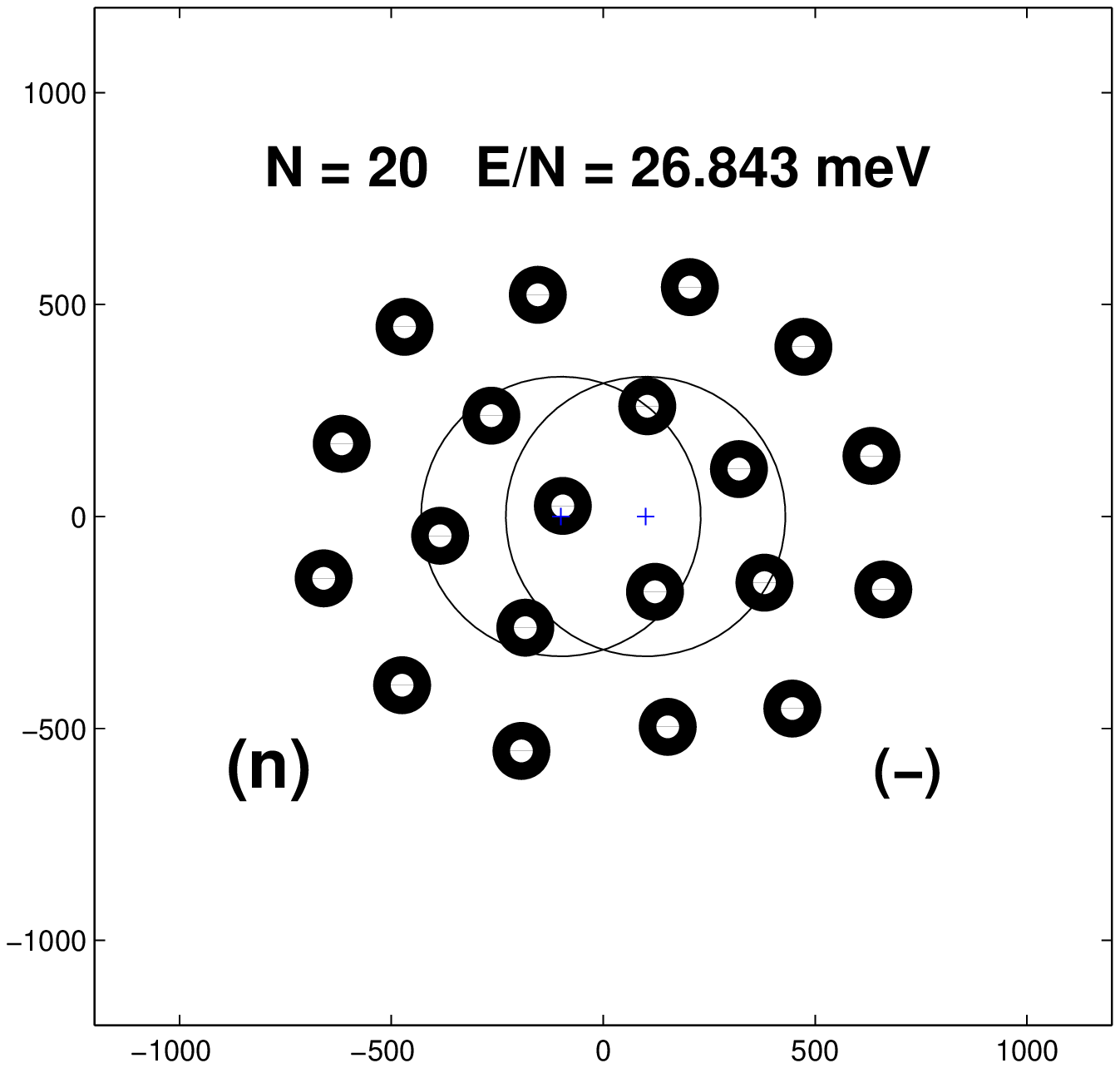} 
&
\includegraphics*[height=25mm]{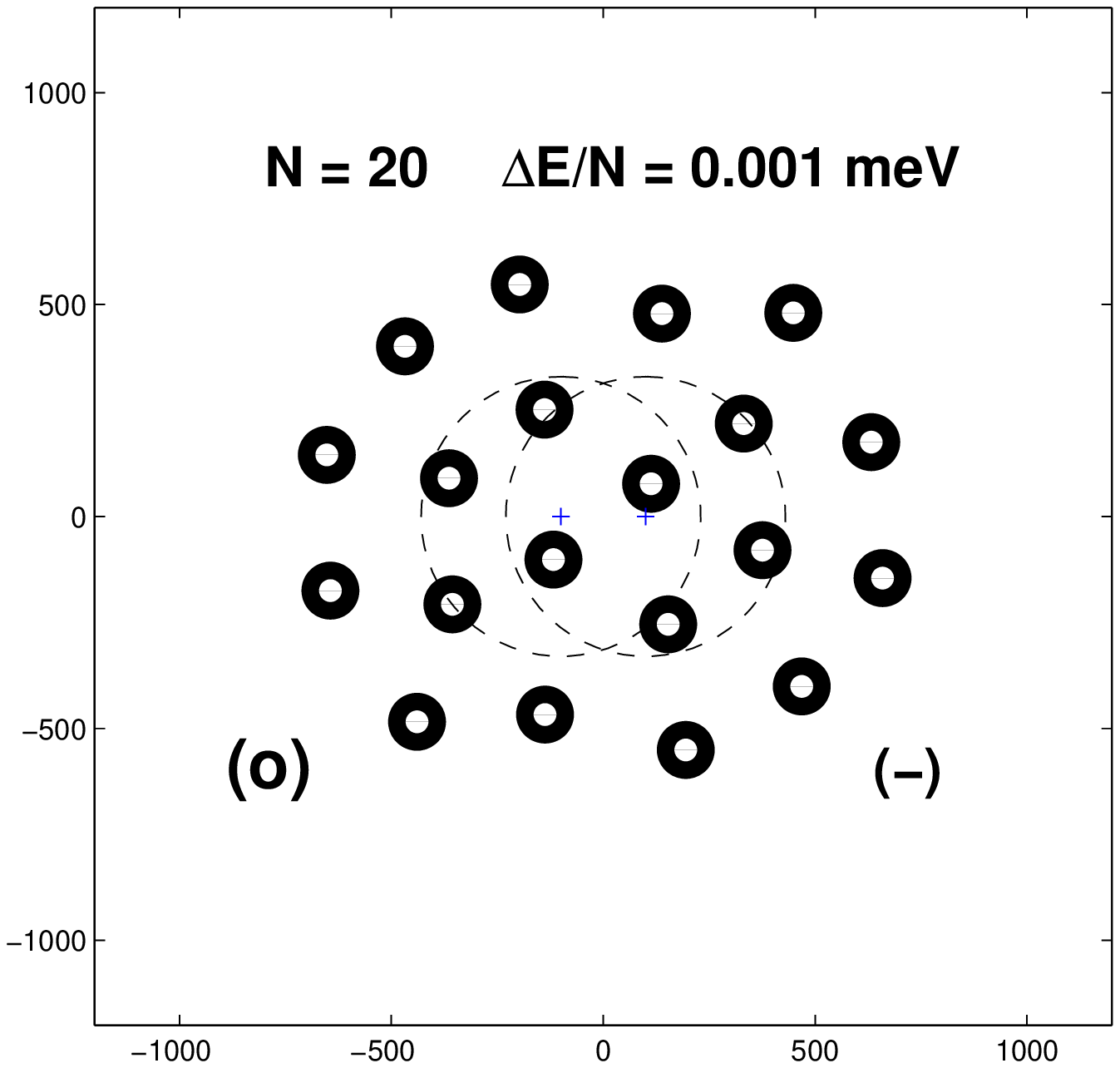} 
\\ \hline
\end{array}$  
\caption{\label{Fig:NN} {\bf(a)} - {\bf(o)}.  Electron configurations of
selected ground and metastable states. The energy per particle ($E/N$) of a
ground state configuration and
the energy difference to the ground state per particle ($\Delta
E/N$) for metastable configuration is given. The configuration is
marked on the lower right corner. 
To make it easier to
distinguish different configurations and the distance between the minima
($d = 0, 200, 600, 1000$ \AA) a
circle with 330 \AA \ radius is 
plotted around each parabolic potential minimum. A dashed circle indicates a
metastable state.} 
\end{figure}

For example, as Table \ref{Tab:N} and Fig \ref{Fig:NN} (a) indicate,
with  
eight particles in the single artificial atom ($d = 0$ \AA) the ground
state is (1,7), one electron in the centre and seven electrons in the
circular shell, and there exist no metastable states.  
At $d = 200$ \AA \ a new ground state has appeared with configuration 
(2,6) (Fig. \ref{Fig:NN} (b)) while (1,7) has changed to a
metastable state (see also Table \ref{Tab:N}).   
At distances $d = 600 $ \AA \  and $d = 1000 $ \AA \
the notation is changed to two-atom configurations and 
for $N = 8$ the ground state is marked with (4),(4), see
Figs. \ref{Fig:NN} (c) and (e). 

The notation for configurations is not always exhaustive. 
The relative orientation of different
shells and the relative orientations of the
configurations of the two atoms
at $d = 600, 1000$ \AA \ do not always become clear from Table
\ref{Tab:N}. 
For example, when either or both atoms are left with four
electrons ($N = 7,8,9$), the orientation of the square(s) relative to
the other atom changes as the distance is increased.     
At smaller distances the position of the square of four
electrons is such that the tips of the squares are in the same line with
the positions of the minima (see Fig. \ref{Fig:NN} (c) for $N = 8$ at $d =
600$ \AA). As the distance is increased the square (or two squares
with $N = 8$) turns onto its side (see Fig. \ref{Fig:NN} (e) for $N = 8$ at
$d = 1000$ \AA).
For $N = 8$ at $d = 600$ \AA \ there also exists
a metastable state where one of the squares is lying on its side
and the other on the tip (Fig. \ref{Fig:NN} (d)). 
Even though we divide
electrons into shells in our notation it does not mean that the shells are
strictly circular even for $d = 0$. This can be seen clearly for $N =
12$ in Fig \ref{Fig:NN} (f), where the outer shell resembles more like
an incomplete triangle with the tips missing. The configuration marked
with (1,5)$^*$ 
in the two-atom configurations in Table \ref{Tab:N} cannot be
identified strictly to (6) but neither to
(1,5). Therefore we choose the notation (1,5)$^*$ to describe the
configuration. The difference between (1,5)$^*$ and (1,5) can be seen
with the (1,5)$^*$,(1,5) metastable state in Fig. \ref{Fig:NN}
(g). The configurations of the ground and metastable states for
$N = 13$ at $d = 200$ \AA \ are marked in the same way in 
the Table, but are different as can be seen in Figs. \ref{Fig:NN}
(h) and (i). 
For the highest energy metastable state for $N = 17, d = 200$ \AA \ the
two-atom notation would have described the configuration better, 
Fig. \ref{Fig:NN} (l). 
One metastable state for $N = 19$ and the ground state and one metastable
state for $N = 20$ at $d 
= 200$ \AA \ could not be described with the shell structure
notation. The configurations are
depicted in Figs. \ref{Fig:NN} (m),(n) and (o), respectively.

\begin{table*}
 \caption{\label{Tab:N} Ground and metastable configurations and
 corresponding 
 energies in meV/particle 
at four studied distances ($d = 0, 200, 600, 1000$
 \AA). The energies of metastable states, $\Delta E/N$, are given as meV/particle
 above the ground state energy. The configurations of electron
 clusters in the two-atom molecule
 are described with concentric shells located around the centre of the
 system at $d = 
 0, 200$ \AA \ and as two separate electron clusters located near
 the minimum of one of the two atoms at $d = 600, 1000$ \AA.   } 
\begin{ruledtabular} 
\begin{tabular}{l|lrr|lrr|lrr|lrr}
 & \multicolumn{3}{c} {$d = 0$ \AA} & \multicolumn{3}{c}{$d = 200$ \AA} & \multicolumn{3}{c}{$d = 600$ \AA} & \multicolumn{3}{c}{$d = 1000$ \AA} \\
N  & $E/N$ & $\Delta E/N$ & config. & $E/N$ & $\Delta E/N$ & config.  & $E/N$ & $\Delta E/N$ & config. & $E/N$ & $\Delta E/N$ & config. \\ 
  & [meV] & [meV] &  & [meV] & [meV] &  & [meV] & [meV] &  & [meV] & [meV] &  \\ \hline
2  & 2.736  &&      (2) & 1.777  &&      (2) & 0.875  &&       (1),(1) & 0.546  &&   (1),(1) \\ \hline
3  & 4.780  &&      (3) &  3.894 &&      (3) & 2.940  &&       (1),(2) & 2.541  &&   (1),(2) \\ \hline
4  & 6.696  &&      (4) &  5.588 &&      (4) & 4.351  &&       (2),(2) & 3.795  &&   (2),(2) \\  \hline
5  & 8.531  &&      (5) &  7.340 &&      (5) & 5.915  &&       (2),(3) & 5.244  &&   (2),(3) \\ 
   && +0.099 &    (1,4) &&        &          &&        &                 &&        &             \\ \hline
6  & 10.231 &&    (1,5) &  8.939 &&      (6) &  7.234 &&       (3),(3) &  6.395 &&   (3),(3) \\ 
   && +0.074 &      (6) && +0.033 &    (1,5) &&        &                 &&        &             \\ \hline
7  & 11.816 &&    (1,6) & 10.459 &&    (1,6) &  8.664 &&       (3),(4) &  7.720 &&   (3),(4) \\ \hline
8  & 13.384 &&    (1,7) & 11.933 &&    (2,6) &  9.932 &&       (4),(4) &  8.850 &&   (4),(4) \\ 
   &&        &          && +0.016 &    (1,7) &&         +0.003  & (4),(4)  & &  & \\ \hline
9  & 14.913 &&    (2,7) & 13.335 &&    (2,7) & 11.246 &&       (4),(5) & 10.097 &&   (4),(5) \\ 
   && +0.022 &    (1,8) &&        &          &&        &                 && & \\ \hline
10 & 16.361 &&    (2,8) & 14.680 &&    (2,8) & 12.441 &&       (5),(5) & 11.195 &&   (5),(5) \\ 
   && +0.012 &    (3,7) &&        &          &&        &                 &&        &             \\ \hline
11 & 17.746 &&    (3,8) & 16.053 &&    (3,8) & 13.701 &&   (5),(1,5)$^*$ & 12.361 &&   (5),(1,5) \\ 
   &&        &          && +0.003 &    (2,9) &&        &                 &&        &             \\ \hline
12 & 19.111 &&    (3,9) & 17.354 &&    (3,9) & 14.873 && (1,5)$^*$,(1,5)$^*$ & 13.434 && (1,5),(1,5) \\ 
   && +0.011 &    (4,8) && +0.004 &    (4,8) && +0.011 & (1,5)$^*$,(1,5) &&        &             \\ \hline
13 & 20.433 &&    (4,9) & 18.624 &&    (4,9) & 16.048 && (1,5)$^*$,(1,6) & 14.511 && (1,5),(1,6) \\ 
   &&        &          && +0.024 &    (4,9) &&        &                    && & \\ \hline
14 & 21.738 &&   (4,10) & 19.854 &&   (4,10) & 17.168 && (1,6),(1,6)     & 15.518 && (1,6),(1,6) \\ 
   && +0.014 &    (5,9) &&        &          &&  &  &&  &  \\ \hline
15 & 23.010 &&   (5,10) & 21.072 &&   (5,10) & 18.302 && (1,6),(1,7) & 16.587 && (1,6),(1,7) \\ 
   && +0.029 &  (1,5,9) && +0.035 &   (5,10) &&        &             && +0.234 & (1,6),(2,6) \\ \hline 
16 & 24.259 && (1,5,10) & 22.271 &&   (6,10) & 19.373 && (1,7),(1,7) & 17.583 && (1,7),(1,7) \\ 
   && +0.009 &   (5,11) && +0.006 &   (5,11) &&        &             && +0.024 & (1,7),(2,6) \\ \hline
17 & 25.473 && (1,6,10) & 23.448 &&   (6,11) & 20.468 && (1,7),(2,7) & 18.611 && (1,7),(2,7) \\ 
   && +0.005 & (1,5,11) && +0.010 & (1,6,10) && +0.006 & (1,7),(1,8) && +0.018 & (1,7),(1,8) \\ 
   &&        &          && +0.016 & (1,5,11) &&        &             && +0.023 & (2,6),(2,7) \\ 
   &&        &          && +0.018 &   (6,11) &&        &             && +0.041 & (2,6),(1,8) \\ \hline
18 & 26.660 && (1,6,11) & 24.597 && (1,6,11) & 21.522 && (1,8),(2,7) & 19.579 && (2,7),(2,7) \\ 
   && +0.026 & (1,7,10) && +0.006 &   (6,12) && +0.001 & (1,8),(2,7) && +0.017 & (2,7),(1,8) \\ \hline
19 & 27.841 && (1,6,12) & 25.728 && (1,6,12) & 22.572 && (2,7),(2,8) & 20.569 && (2,7),(2,8) \\ 
   && +0.003 & (1,7,11) && +0.004 & (1,7,11) && +0.001 & (1,8),(2,8) && +0.016 & (1,8),(2,8) \\ 
   &&        &          && +0.016 &    -     &&        &             &&        &             \\ \hline
20 & 29.000 && (1,7,12) & 26.843 &&    -     & 23.583 && (2,8),(2,8) & 21.585 && (2,8),(2,8) \\ 
   && +0.024 & (1,6,13) && +0.001 &    -     &&        &             &&        &             \\ 
   &&        &          && +0.003 & (2,7,11) &&        &             &&        &             \\ 
 \end{tabular}
\end{ruledtabular}
\end{table*}

\begin{figure*}
\begin{center}
   \includegraphics*[height=60mm]{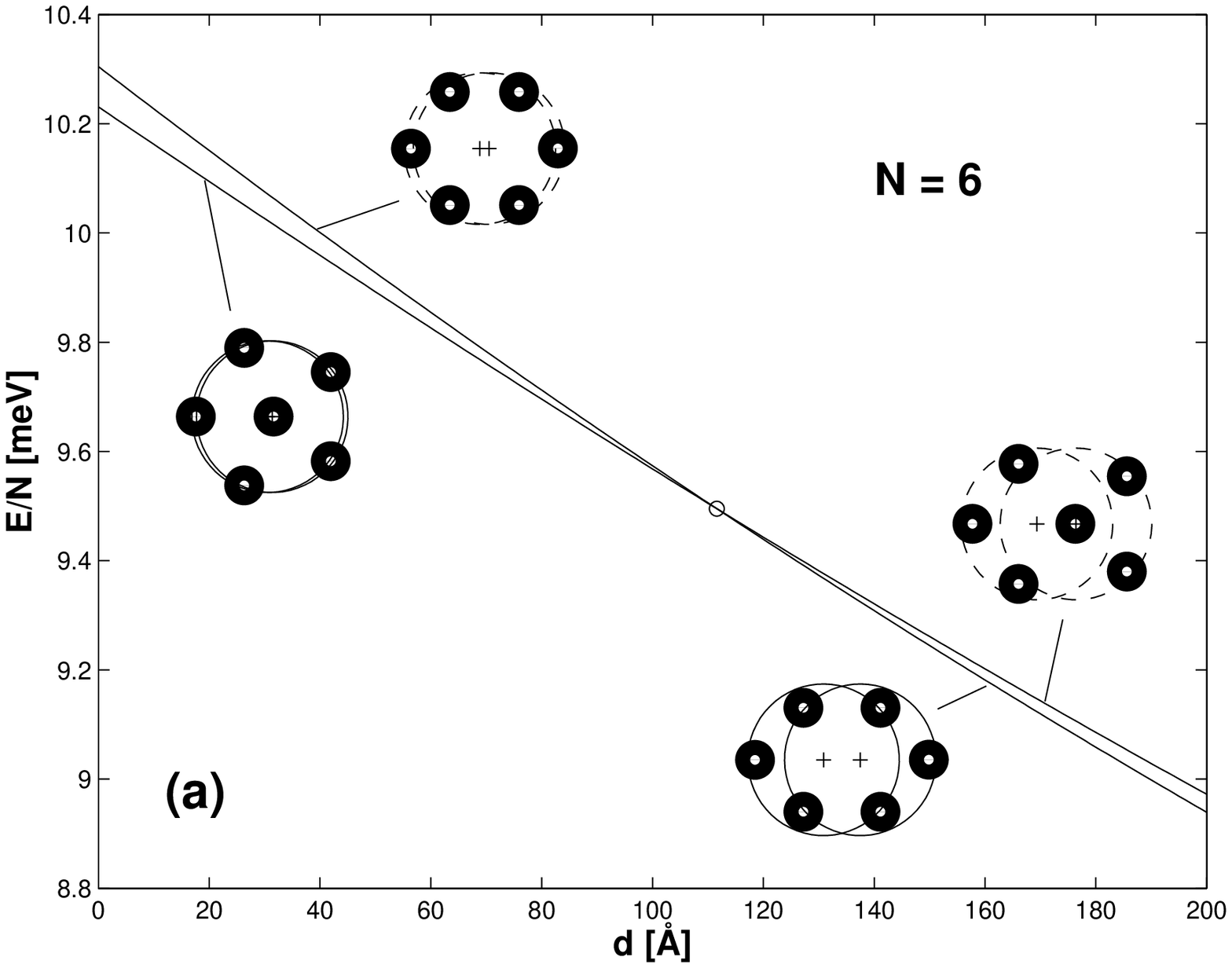} 
   \includegraphics*[height=60mm]{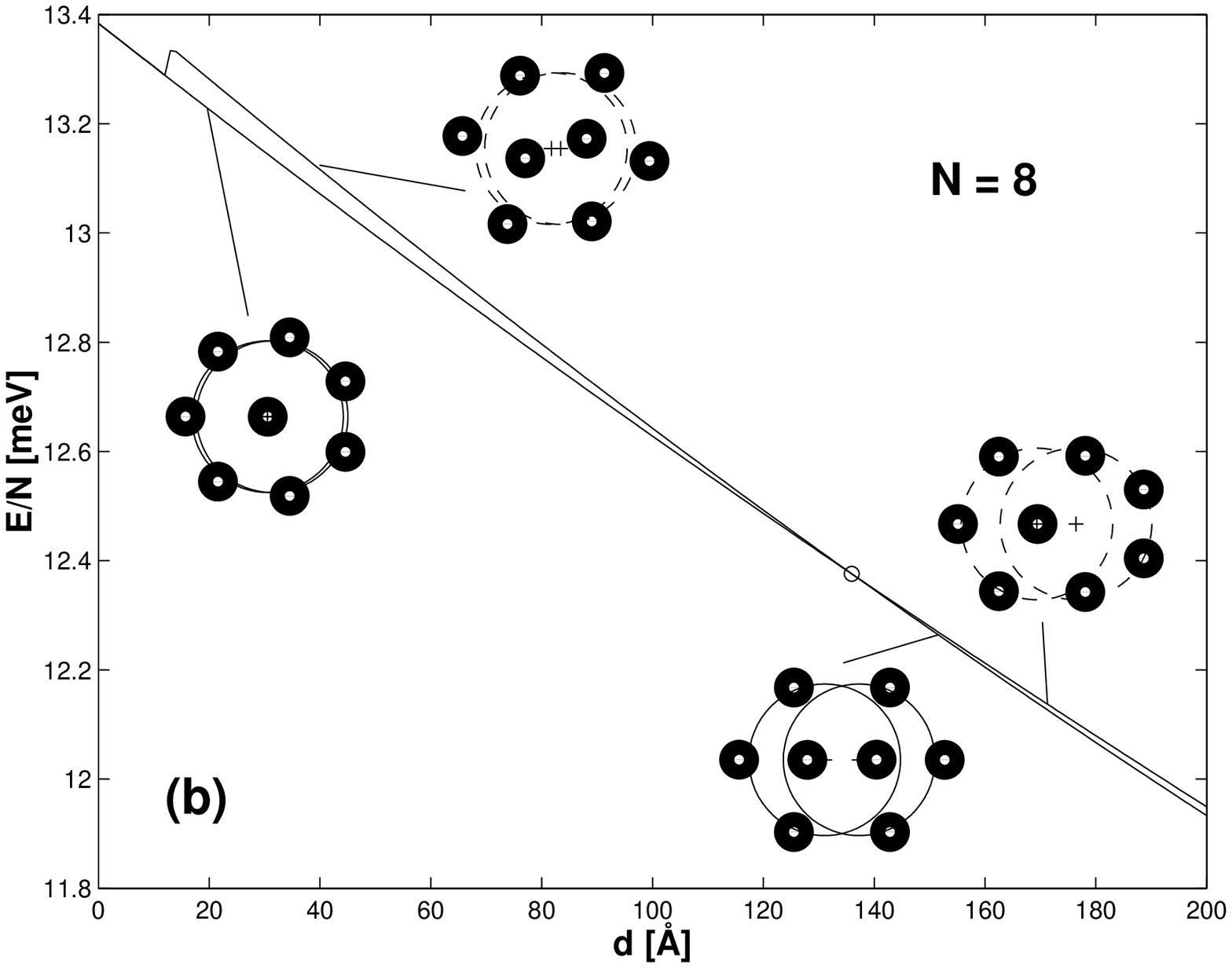} 
   \includegraphics*[height=60mm]{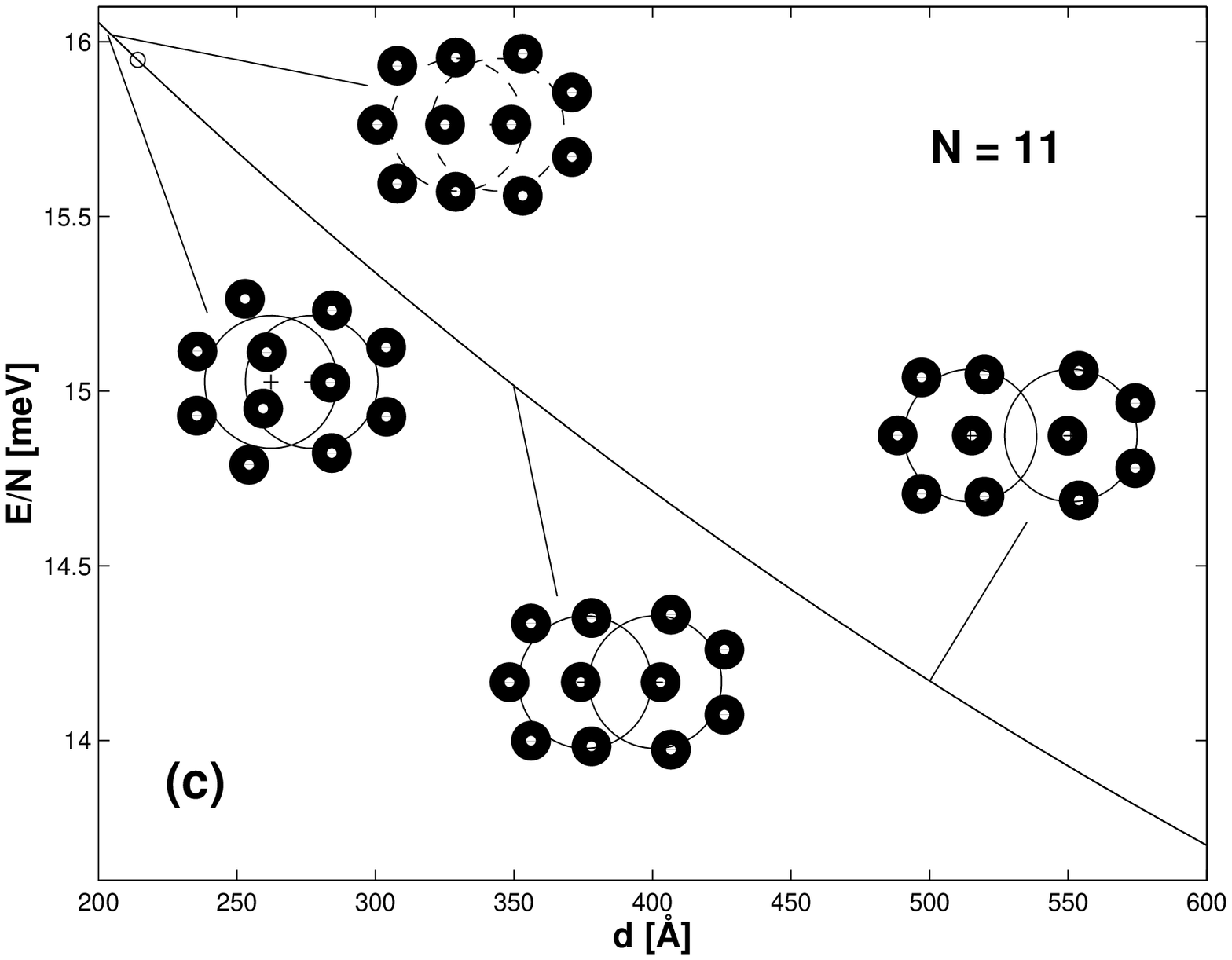} 
   \includegraphics*[height=60mm]{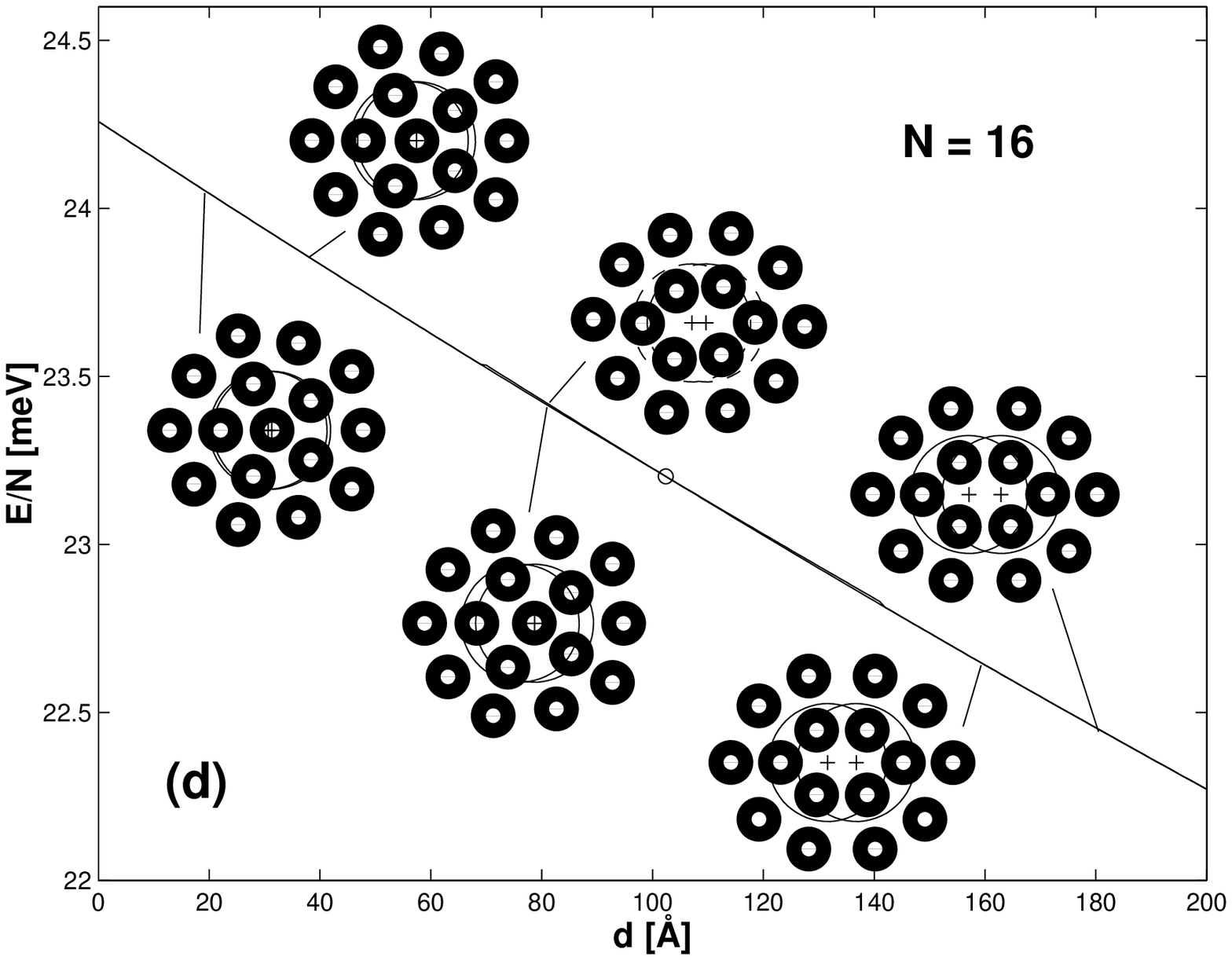}
   \includegraphics*[height=60mm]{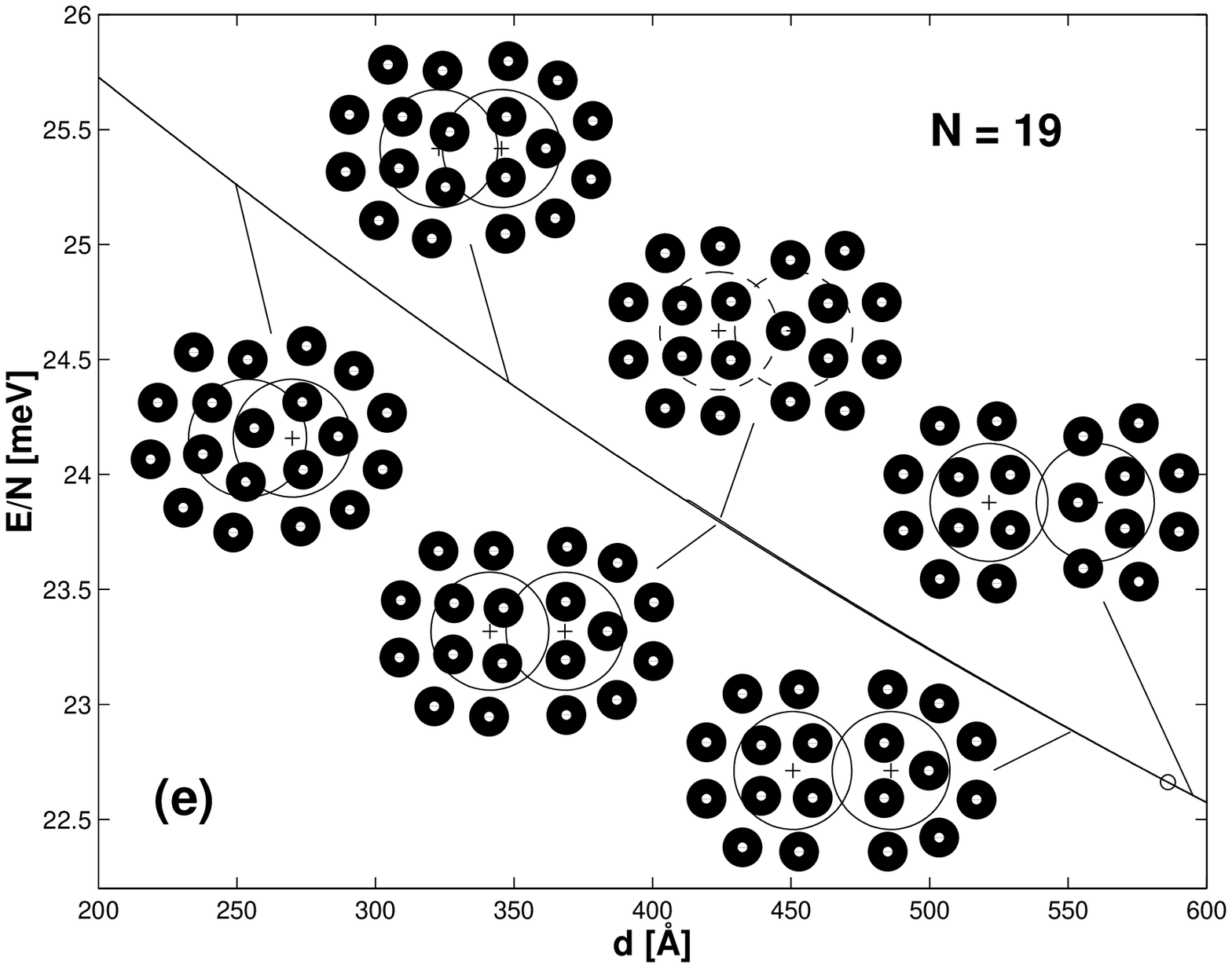} 
\end{center}
\caption{\label{Fig:N6N8N16N19} {\bf (a)} - {\bf (e)}. Energy per
particle as a function of distance for $N = 6, 8, 11, 16, 19$. The small
circle indicates the transition point. To make it easier to
distinguish different configurations and the distance between the minima a
circle with 330 \AA \ radius is 
plotted around each parabolic minimum. A dashed circle indicates a
metastable state.  
}
\end{figure*}

As the distance between the atoms is increased it is not always clear
whether the electrons just follow the two atoms drawn apart and
continuously change to two separated atoms.
Sometimes metastable states change to a ground state and the
ground state to a metastable state as the distance between the atoms is
increased. The clearest example can be seen in Table \ref{Tab:N} for
six electrons between $d = 0$ and $d = 200 $ \AA. At $d = 0$ the (1,5)
configuration is the ground state and (6) the metastable state. 
At $d = 200$ \AA \
 it is the other way around: (6) is the ground and (1,5) a
metastable state. The energy as a function of distance for two
alternative configurations is shown in Fig.\ref{Fig:N6N8N16N19} (a).
The transition point, marked with a small circle, is at $d = 111.6 $ \AA. 
The transition is continuous with respect to energy
as a function of distance, 
but the curvature of the $E(d)$-curve is different 
and therefore the 
first derivative of energy with respect to $d$ is discontinuous. This
is a first-order discontinuous structural transition in the electron
configuration. Hereafter, by discontinuous structural 
transitions we mean the qualitative change in the ground state
electron configuration which is discontinuous with respect to $\partial
E/\partial d$ at the transition point.       

In addition to $N = 6$, for $N = 8, 11, 16$ and $19$ one qualitative change
in the electron configuration is observed as a function of $d$. For $N
= 8$ at $d = 135.9$ \AA \ the electron configuration changes from (1,7) to
(2,6), see Fig \ref{Fig:N6N8N16N19} (b). Notice that the (2,6)
configuration is not stable at the limit of 
one atom and becomes unstable approximately below $d = 17$ \AA. 
For $N = 11$ there exists one metastable state, (2,9), at $d = 200$ \AA,  which
at $214.2$ \AA \ changes to a ground state
as is depicted in Fig. \ref{Fig:N6N8N16N19} (c). 
With
$N=16$ the configuration changes from (1,5,10) to (6,10) at $d = 102.4
$ \AA. The energy differences between ground and metastable states are
small and metastable states exist only in the proximity of
the transition point 
(Fig. \ref{Fig:N6N8N16N19} (d)). At $d = 80 $ \AA \ \ the (1,5,10)
configuration which is plotted below the energy curve is still the
ground state and the (6,10) configuration plotted above the energy
curve has appeared as a metastable state. 
For $N = 19$ 
Fig. \ref{Fig:N6N8N16N19} (e) shows how the (1,6,12)
configuration, which is the ground state at $d = 200 $ \AA, has
changed to a rather unsymmetric configuration at $d 
= 250$ \AA. This unsymmetric configuration changes continuously to
(2,8),(1,8) (lower middle plot in (e)) which 
at $586.0 $ \AA \  changes discontinuously to (2,8),(2,7)
configuration. The (2,8),(2,7) configuration appears approximately at
$415$ \AA \ as a metastable state. The (1,8),(2,8) persists as a
metastable state to the greatest studied distance of $1000 $ \AA \ as
is also indicated in Table \ref{Tab:N}. The other metastable states at
$d = 0$ and $200$ \AA \ do not change to a ground state. 

\begin{figure}
   \includegraphics*[height=60mm]{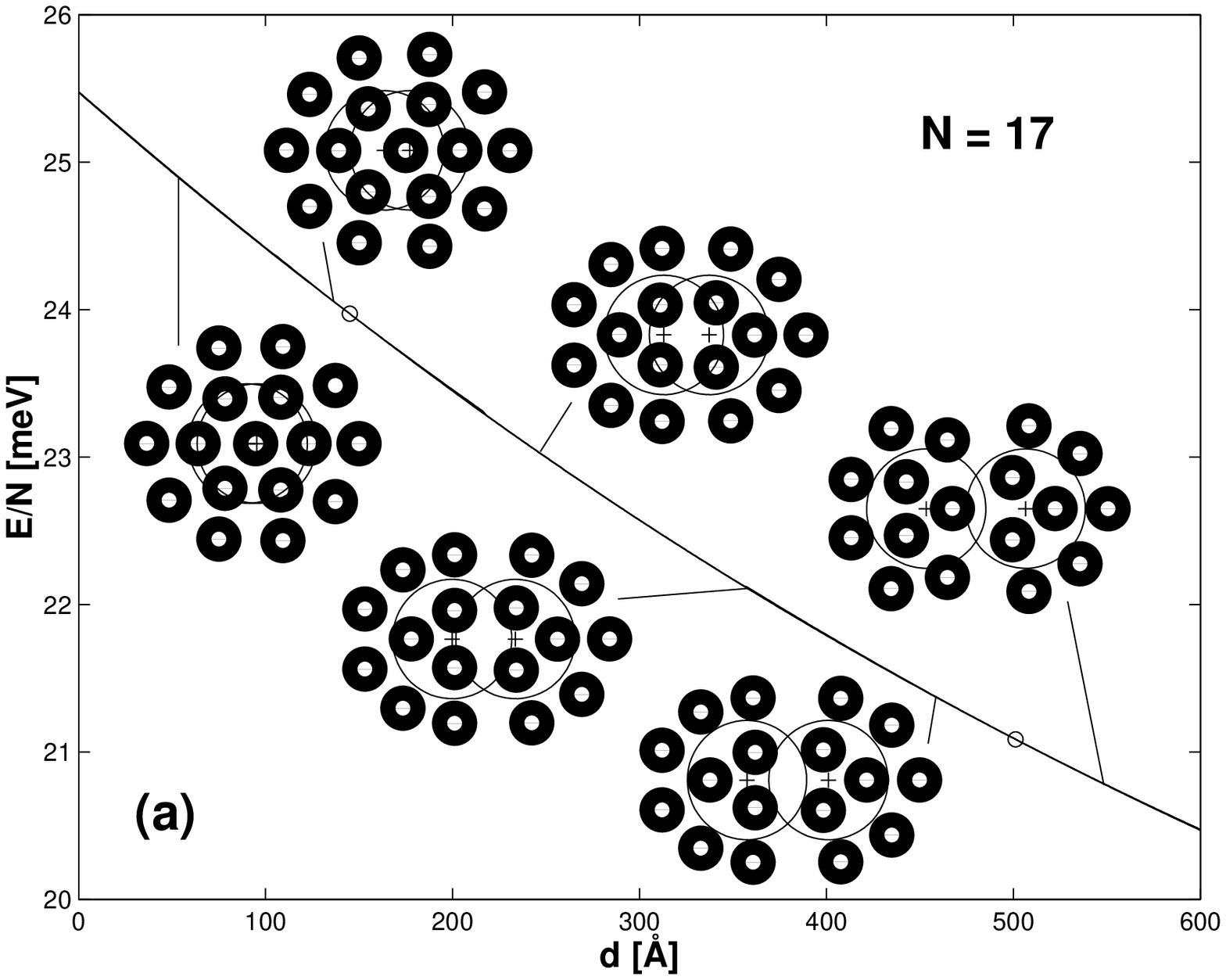}
   \includegraphics*[height=60mm]{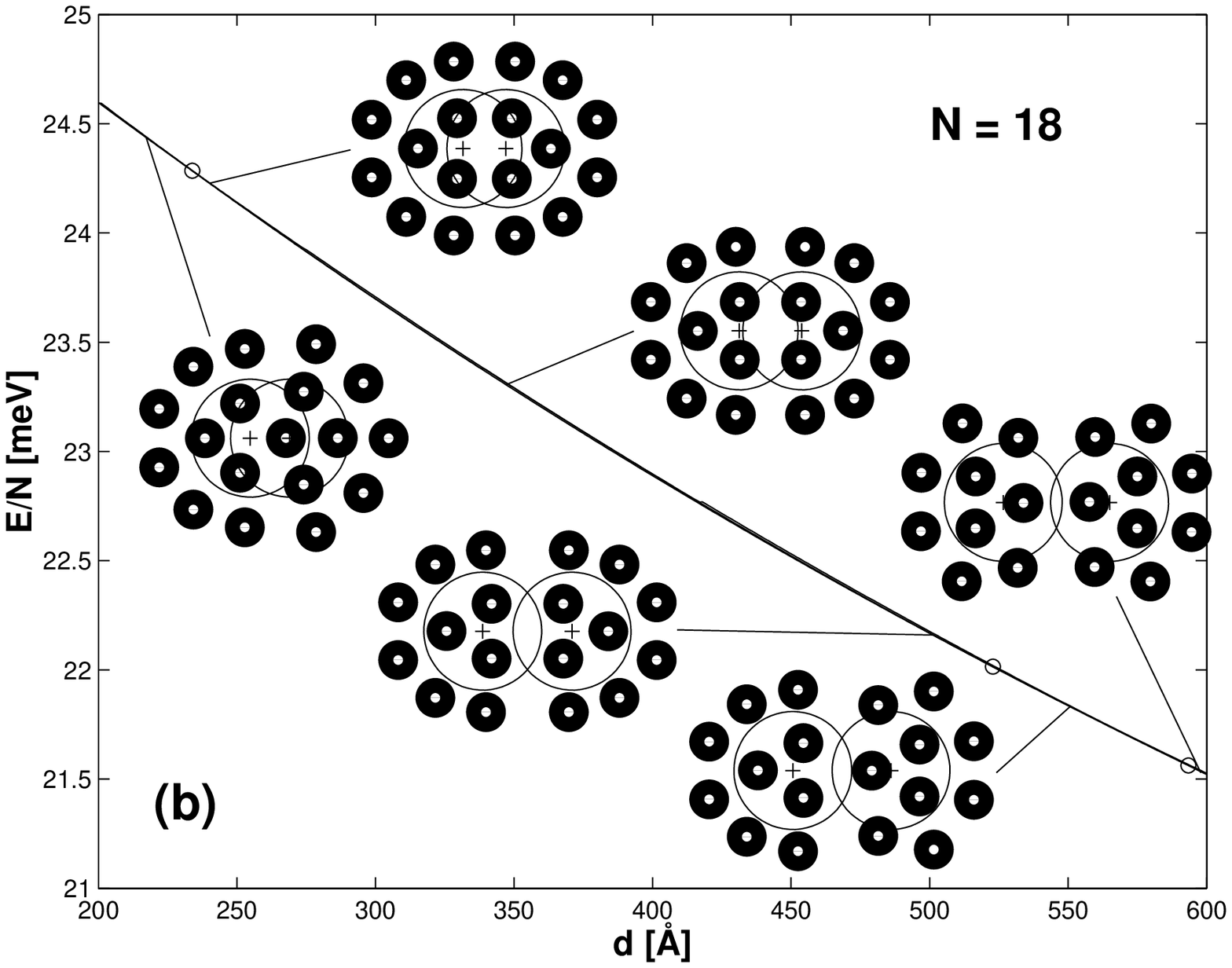}
\caption{\label{Fig:N17N18} Energy as a function of distance for $N =
17$ and $N = 18$ with the ground state configurations along the energy
curve. The small circles indicate the discontinuous structural transition
points.}
\end{figure}

More than one discontinuous transformation in the electron
configurations is found for $N = 17$ and
$18$, see Figs. \ref{Fig:N17N18} (a) and (b). For $N = 17$ the (1,6,10)
changes to (6,11) at $d = 145.0$  
\AA.
 The (6,11) configuration parts to (1,7),(1,8) two-atom configuration,
which is the ground state up to $d = 501.1$ \AA, where the (2,7) configuration
of the other atom becomes more stable than (1,8), thus
(1,7),(1,8)$\rightarrow$(1,7),(2,7).     
The (1,7),(2,7) configuration persists as a ground
state to greater distances. Qualitatively similar transformations are seen for
$N = 18$ as for $N = 17$. First the centred cluster (1,6,11) changes to
an open configuration (6,12) at $d = 233.9$ \AA. The open configuration
follows the 
separation of atoms adopting the configuration
(1,8),(1,8), where as in $N = 17$, the (2,7) becomes more stable than
(1,8). However, we now see two discontinuous transformations. 
The first at $d = 522.8$ \AA \
when (1,8),(1,8)$\rightarrow$(2,7),(1,8)
and the second at $d = 593.4$ \AA \ when
(2,7),(1,8)$\rightarrow$(2,7),(2,7).

\begin{figure}
   \includegraphics*[height=60mm]{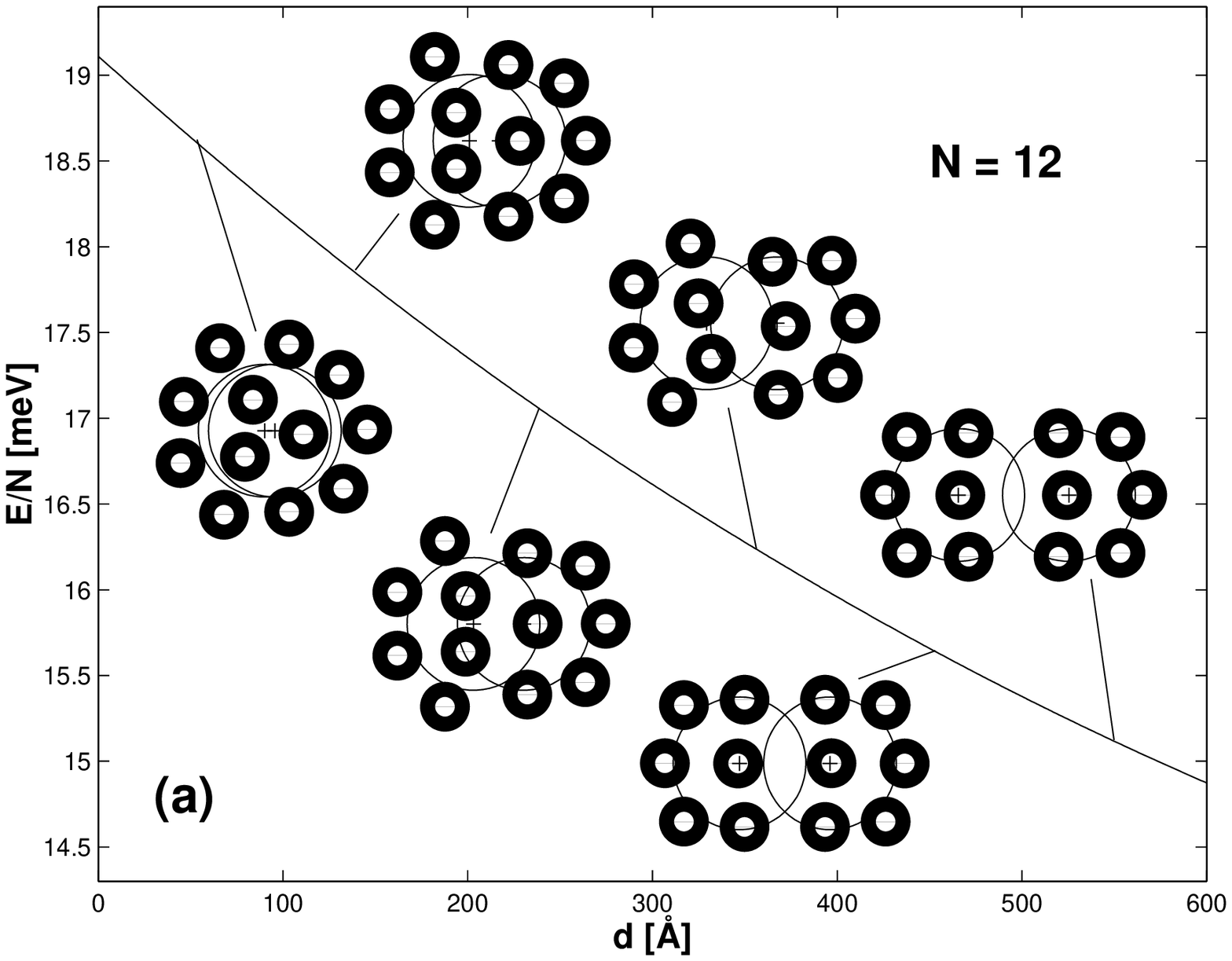}
   \includegraphics*[height=60mm]{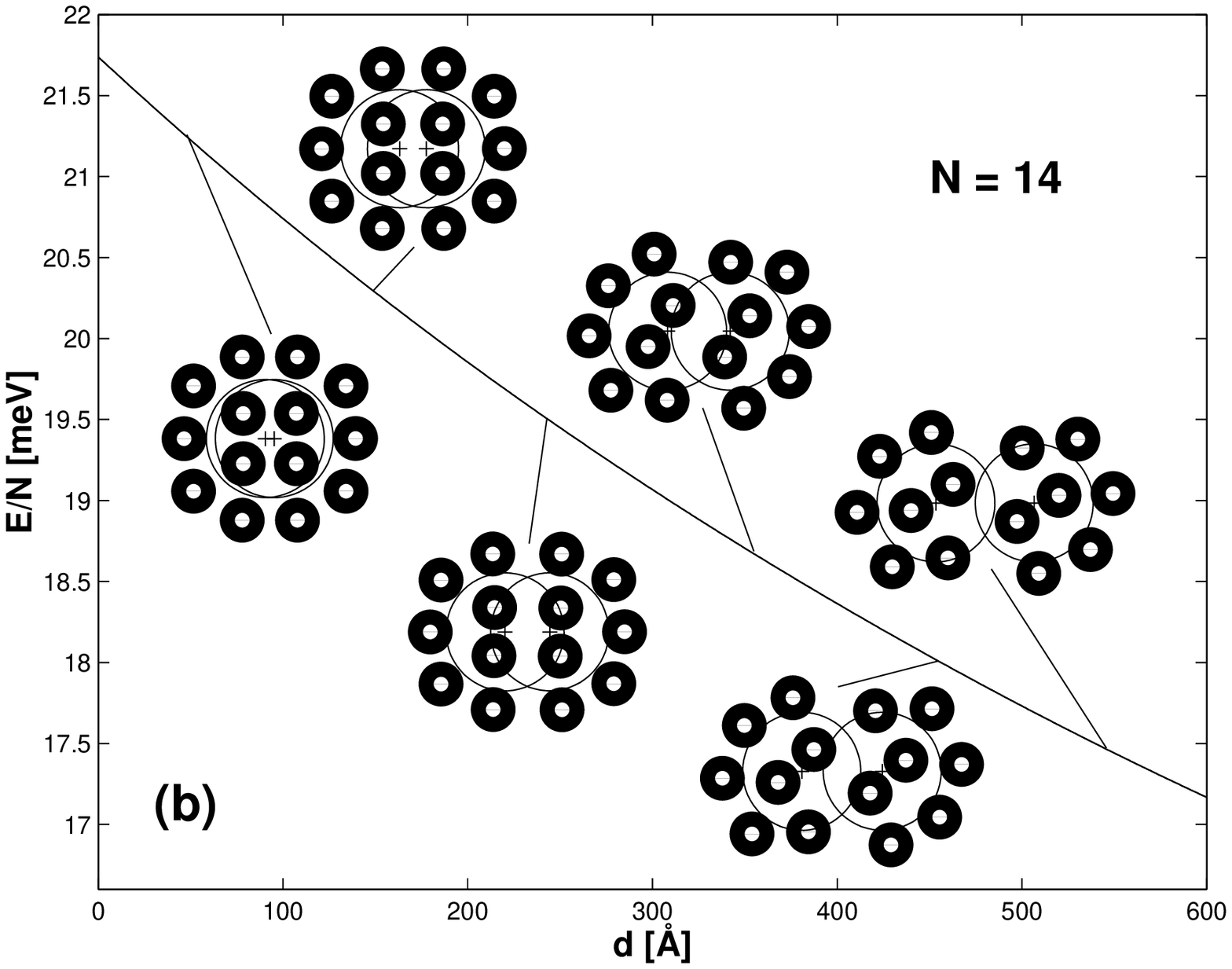}
   \includegraphics*[height=60mm]{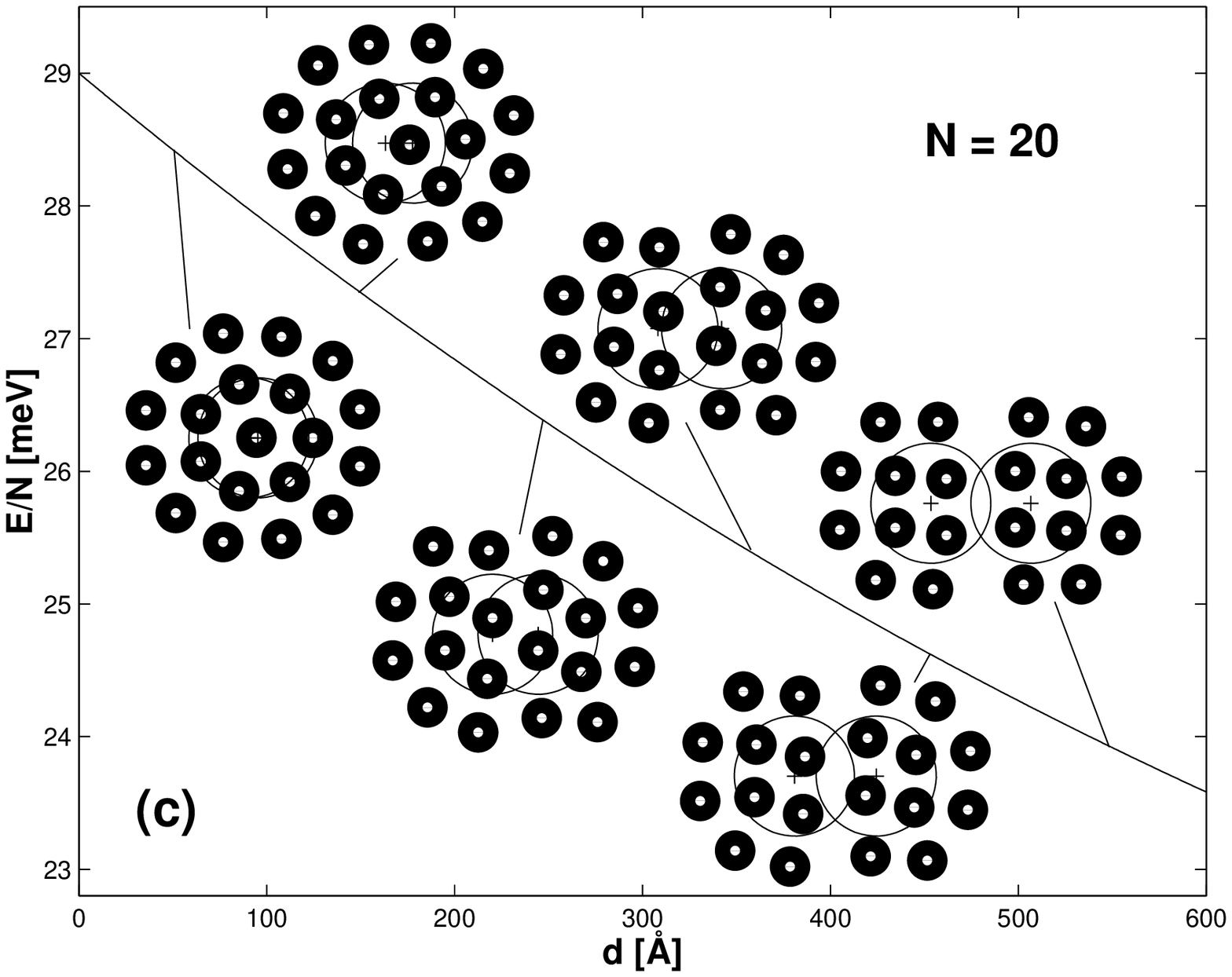}
\caption{\label{Fig:N12N14N20} {\bf (a)} - {\bf (c)} Ground state
electron configurations along $E(d)$ curve for $N = 12, 14$ and $20$.} 
\end{figure}

For other electron numbers besides the reported $N = 6,8,16,17,18,19$
we do not observe discontinuous structural transitions in the electron
configuration as the distance between the two atoms is increased. A few
examples of continuous electron configuration changes are shown in
Fig. \ref{Fig:N12N14N20}. For $N = 12$ the (3,9) configuration transforms
continuously to resemble the (6),(1,5)$^*$ two-atom configuration. Between
$200$ \AA \ and $450$ \AA \ the row of electrons pushes itself forward
when the atoms move apart, resulting in the symmetric configuration
(1,5)$^*$,(1,5)$^*$. For $N = 14$ the electron configuration follows the
separation of atoms in a symmetrical form, but after $d = 250$ \AA \ both sides
start to twist towards the (1,6),(1,6) two-atom configuration. 
For $N = 20$ the transformation is hard to describe, but it is
continuous. The three distinct metastable states at $d = 0$ and $200$
\AA, seen in Table \ref{Tab:N}, never change to a ground state and vanish
at other distances. 

\begin{figure}
   \includegraphics*[height=60mm]{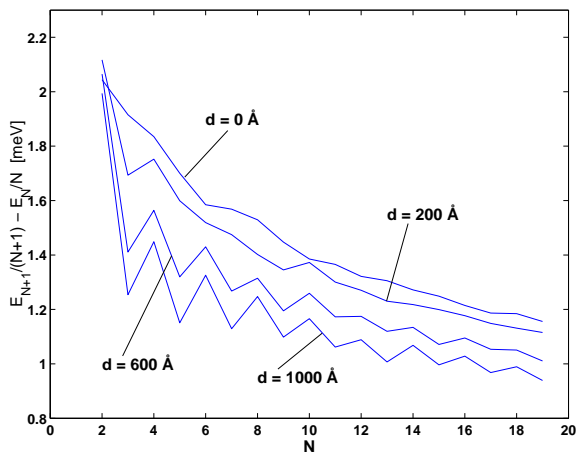}
\caption{\label{Fig:dmyy} Change in the chemical potential ($E_{N+1} /
(N+1) - E_N / N$) at $d = 0, 200, 600$ and $1000$ \AA.}
\end{figure}

The changes in energy per particle of the ground states at the four studied
distances as a function of $N$ are shown in Fig. \ref{Fig:dmyy}. At
$d =$ 0 there are small troughs at $N = 3, 6, 10$ and $17$, at
adding the  
fourth, seventh, eleventh and eighteenth particle. Moving to greater distances
between the atoms, the change in the chemical potential is clearly
peaked. Going to an odd number of particles increases 
the chemical potential much more than going to an even number of
particles. Interesting is the
intermediate distance of $d =$ 200 \AA \ where this trend is observed
for $N = 2,3,4$ and $9,10,11$, but otherwise the curve shows
no clear structure and does not strictly follow the shape of the $d
= 0$ \AA \ curve either.

\section{\label{sec:discussion}Discussion}

At $d = 0$ (single artificial atom) our results are in agreement with
other Monte Carlo (MC) and molecular dynamics (MD) 
studies with parabolic confinement and pure Coulomb interaction.
\cite{Bolton_Superlatt,BedanovPeeters,Schweigert,Ying-Ju,Date,Kong}
However, for $N =
17$ Bolton \etal~\cite{Bolton_Superlatt} obtain the (1,5,11)
configuration instead of 
(1,6,10) which our and other
calculations\cite{BedanovPeeters,Schweigert,Ying-Ju,Kong} predict. 
Ref. \onlinecite{Bolton_Superlatt} may contain
an error since in a later work by Bolton\cite{Bolton's_thesis} the
configuration for $N = 17$ 
was reported to be (1,6,10). There is also a difference for $N = 21$ in
Ref. \onlinecite{Bolton_Superlatt} which was later
corrected.\cite{Bolton's_thesis}   
 
Besides calculating the ground state configurations 
Kong \etal~\cite{Kong} also examined metastable states for $N =
1 - 40$. Our results are in agreement (we calculated
configurations only for $N \leq 20$) both in ground and 
metastable states except that for $N =
18$ Kong \etal
found two metastable states whereas we see only one. In addition to
the (1,6,11) and (1,7,10) they also obtain (6,12) as a metastable
state. 
We repeated the simulation with 3000 independent
test runs, but were still unable to find the (6,12) configuration.

We can conclude that different calculations for the $r^2$
confinement and $1/r$ interaction potential are in good general
agreement. The few experiments 
on charged particles trapped in 2D as well as  calculations with different
forms of interaction and confinement potentials reveal also different
configurations for the
cluster patterns. The interaction between the particles could be logarithmic,
which is the case with infinite charged lines moving in 2D (vortex
lines etc.) or 
perhaps Yukawa type with a strong but short-range repulsion (screened
Coulomb interaction). The form of
the confinement is usually chosen to be parabolic (Lai and I~\cite{Ying-Ju} 
tested also 
a steeper confinement with $r^4$ contribution). 
However, for the question whether the potential in
experiments with clusters is parabolic there is no clear
answer. Therefore it is not surprising that the experiments and also
calculations with different functional forms of interaction and
confinement result in different cluster patterns.    

Lai and I~\cite{Ying-Ju} calculated and
summarised the configuration patterns with different interactions and
tested also $r^4$ contribution to the confinement  
and compared the
results to dust particle experiments.~\cite{dustparticles} Saint Jean
\etal~\cite{exp_classical}  
measured the configurations with electrostatically interacting charged
balls of millimeter size moving on a plane conductor. They made a
comparison with simulations and quite surprisingly found the best agreement
with a relatively old simulation with vortex lines in a
superfluid\cite{Campbell} with 
logarithmic interaction, which again was not in agreement with the
dust particle 
experiments\cite{dustparticles} nor with the purely
logarithmic interaction 
used by Lai and I.~\cite{Ying-Ju}  Despite the differences there are some 
particle numbers where the configuration seems to be the same
regardless of the
experiment or functional form of the interaction or confinement. These
particle numbers are $N = 3,4,5,7,10,12,14,19$.

Partoens \etal \cite{PartoensClassicalPos} examined the ground states of
even number of 
classical electrons evenly distributed in two {\it vertically} coupled
artificial atoms as a function of the distance between the
atoms. As in our study of {\it laterally} coupled atoms discontinuous
transitions between configurations occur as a function of the distance
$d$. The difference is that for vertically coupled atoms one can see
intuitively that transitions should occur between $d = 0$ and $d
\rightarrow \infty$ whereas in laterally coupled atoms the configurations
can be pulled apart with some $N$ without qualitative (discontinuous)
changes in the electron 
configurations. For vertically coupled atoms  
discontinuous transitions (first order with respect to energy) in electron
configurations are observed for all even $N \leq 20$ whereas for
laterally coupled atoms we see also purely continuous
changes as $d \rightarrow \infty$. 

To summarise, we have calculated ground and metastable
state configurations of 
classical point charges confined in two dimensions with two laterally
coupled parabolic potential wells. 
Ground and metastable electron configurations were studied
as a function of the distance between the atoms and discontinuous (in
$\partial E / \partial 
d$) transitions in the ground state configurations were observed 
for particle numbers $N = 6,8,11,16,17,18,19$.      
The configurations of purely classical electrons in laterally coupled
two-minima potential have
an interesting and complex spectrum as the distance
between the minima is changed. 

\begin{acknowledgments}

This work has been supported by the Academy of Finland through its
Centers of Excellence Program (2000-2005). 
  
\end{acknowledgments}


\end{document}